\documentclass[aps,prx,twocolumn,superscriptaddress,footinbib,floatfix]{revtex4-2}
\usepackage{graphicx}
\usepackage{indentfirst}
\usepackage{braket}
\usepackage{float}
\usepackage{amsmath}
\usepackage{physics}
\usepackage{amssymb}
\usepackage{CJK}
\usepackage{esint}
\usepackage{color}
\usepackage[T1]{fontenc}
\usepackage{subfigure}
\usepackage{amsfonts}
\usepackage{footmisc}
\usepackage{scrextend}
\usepackage{multirow}
\usepackage[outdir=./]{epstopdf}

\usepackage[english]{babel}
\usepackage{url}
\usepackage{comment}
\usepackage{array}
\usepackage{subfiles}
\usepackage{tikz}
\usetikzlibrary{shapes,arrows}
\usepackage{mathrsfs}
\usepackage[mathscr]{eucal}

\usepackage[hyperfootnotes=false]{hyperref}
\usepackage{cleveref}
\definecolor{darkblue}{rgb}{0,0,0.5}
\hypersetup{
colorlinks=true,
linkcolor=black,
filecolor=blue,
citecolor=darkblue,
urlcolor=black,
}

\def\be{\begin{equation}}
\def\ee{\end{equation}}
\def\ba{\begin{eqnarray}}
\def\ea{\end{eqnarray}}
\def\bal{\begin{equation}\begin{aligned}}
\def\eal{\end{aligned}\end{equation}}

\def\bp{\begin{pmatrix}}
\def\ep{\end{pmatrix}}

\usepackage{bm}

\newcommand{\1}{^{(1)}}

\usepackage{pict2e,picture,graphicx}
\makeatletter
\DeclareRobustCommand{\Arrow}[1][]{%
\check@mathfonts
\if\relax\detokenize{#1}\relax
\settowidth{\dimen@}{$\m@th\rightarrow$}%
\else
\setlength{\dimen@}{#1}%
\fi
\sbox\z@{\usefont{U}{lasy}{m}{n}\symbol{41}}%
\begin{picture}(\dimen@,\ht\z@)
\roundcap
\put(\dimexpr\dimen@-.7\wd\z@,0){\usebox\z@}
\put(0,\fontdimen22\textfont2){\line(1,0){\dimen@}}
\end{picture}%
}
\makeatother

\newcommand{\QZ}[1]{{{\textcolor{black}{#1}}}}

\begin{document}

\title{Uncovering Quantum Many-body Scars with Quantum Machine Learning}

\author{Jia-Jin Feng}
\email{jiajinfe@usc.edu}
\affiliation{Ming Hsieh Department of Electrical and Computer Engineering, University of Southern California, Los Angeles, California 90089, USA}

\author{Bingzhi Zhang}
\affiliation{Department of Physics and Astronomy, University of Southern California, Los
Angeles, California 90089, USA}
\affiliation{Ming Hsieh Department of Electrical and Computer Engineering, University of Southern California, Los Angeles, California 90089, USA}

\author{Zhi-Cheng Yang}
\email{zcyang19@pku.edu.cn}
\affiliation{School of Physics, Peking University, Beijing 100871, China}
\affiliation{Center for High Energy Physics, Peking University, Beijing 100871, China}

\author{Quntao Zhuang}
\email{qzhuang@usc.edu}
\affiliation{Department of Physics and Astronomy, University of Southern California, Los Angeles, California 90089, USA}
\affiliation{Ming Hsieh Department of Electrical and Computer Engineering, University of Southern California, Los Angeles, California 90089, USA}

\begin{abstract}
Quantum many-body scars are rare eigenstates hidden within the chaotic spectra of many-body systems, representing a weak violation of the eigenstate thermalization hypothesis (ETH). Identifying these scars, as well as other non-thermal states in complex quantum systems, remains a significant challenge.  Besides exact scar states, the nature of other non-thermal states lacking simple analytical characterization remains an open question.
In this study, we employ tools from quantum machine learning---specifically, (enhanced) quantum convolutional neural networks (QCNNs), to explore hidden non-thermal states in chaotic many-body systems. Our simulations demonstrate that QCNNs achieve over 99\% single-shot measurement accuracy in identifying all known scars. Furthermore, we successfully identify new non-thermal states in models such as the xorX model, the PXP model, and the far-coupling Su-Schrieffer-Heeger model.
In the xorX model, some of these non-thermal states can be approximately described as spin-wave modes of specific quasiparticles. We further develop effective tight-binding Hamiltonians within the quasiparticle subspace to capture key features of these many-body eigenstates. Finally, we validate the performance of QCNNs on IBM quantum devices, achieving single-shot measurement accuracy exceeding 63\% under real-world noise and errors, with the aid of error mitigation techniques. Our results underscore the potential of QCNNs to uncover hidden non-thermal states in quantum many-body systems.
\end{abstract}
\maketitle

\section{INTRODUCTION}

Chaos is a fascinating phenomenon in both classical and quantum systems. In chaotic systems, small perturbations can lead to vastly different dynamical trajectories due to the extreme sensitivity to initial conditions, making long-term predictions highly challenging. The ergodicity and mixing properties of chaotic dynamics underpin the principles of statistical physics and thermodynamics \cite{von2010proof,reimann2008foundation,Zhang2016}. Occasionally, amidst the chaotic spectrum, there are sometimes integrable, periodic trajectories known as ``scars", where the system temporarily exhibits regular behavior. In contrast, integrable systems are more straightforward to analyze, as their motions are fully predictable and often display periodic dynamics. Besides fully chaotic and integrable dynamics, there exists a class of pseudo-integrable dynamics, where certain chaotic characteristics are present despite some Lyapunov exponents being zero~\cite{richens1981pseudointegrable,dragovic2015periods}. These dynamics blend regular and complex behavior, sitting between full integrability and complete chaos.

In quantum many-body systems, chaos reveals itself through the properties of eigenstates, even though the dynamics are governed by the intrinsically linear Schrödinger equation. The eigenstate thermalization hypothesis (ETH)~\cite{srednicki1994chaos,deutsch1991,rigol2008thermalization,popescu2006entanglement,Kim2014} assumes that in chaotic systems, the expectation values of local observables appear thermal, even within a single eigenstate. These eigenstates exhibit volume-law scaling of entanglement entropy.  When the system evolves from a generic initial state, the phases of the amplitudes on different eigenstates `randomize' rapidly in the unitary evolution, causing the expectation values of local observables to equilibrate to the diagonal ensemble~\cite{short2012quantum}.

In certain quantum many-body systems, the eigenstate thermalization hypothesis (ETH) can be weakly violated, allowing non-chaotic eigenstates to persist within an otherwise chaotic spectrum. These rare eigenstates, known as quantum many-body scars (QMBSs), represent a vanishing fraction of the entire eigenstates~\cite{Serbyn2021,moudgalya2022quantum,Hang2023,Evrard2024}.
In contrast to chaotic eigenstates, QMBSs offer an intriguing window into non-thermal behavior, posing a challenge for traditional analytical methods.
Experimental studies with a 51-atom quantum simulator have observed persistent revivals over long timescales~\cite{bernien2017probing}, attributed to an evenly spaced tower of QMBSs~\cite{Serbyn2021,moudgalya2022quantum}.  These states are characterized by sub-volume-law entanglement entropy, setting them apart from the typical chaotic eigenstates \cite{Iadecola2020}.
The strict definition of QMBS remains an open problem, complicated by the complex nature of many-body systems. Identifying atypical eigenstates, which could be candidates for QMBSs, is a challenging task.  It is pivotal for understanding ergodicity breaking in quantum systems.
Some of these states may be linked to pseudo-integrability due to finite energy and Hilbert space \cite{Wang2021, Eswarathasan2017,Kong2024}. The weak integrability breaking can exhibit longer thermalization times in quantum many body systems \cite{Surace2023,ivanov2024volumepxp}.

Quantum machine learning (QML) offers powerful tools to study quantum-many-body physics. \QZ{Among these, quantum convolutional neural networks (QCNNs) are designed to detect patterns in quantum systems and have demonstrated advantages over classical convolutional neural networks when processing classical input data~\cite{Mordacci2024}.} QCNNs have proven particularly effective in classifying different phases of matter in quantum spin systems, as demonstrated both theoretically~\cite{Lukin2019, maccormack2022branching} and experimentally~\cite{herrmann2022realizing}. In particular, QCNNs enable model-independent learning, where the phase boundary can be predicted from synthetic fixed-point states associated with each phase individually~\cite{liu2023model}. However, identifying quantum many-body scars (QMBSs) presents a greater challenge than identifying phases due to the scarcity of data, as QMBSs constitute only a vanishingly small fraction of the eigenstates.
Recent studies~\cite{Tomasz2022,cao2024} have applied unsupervised learning to investigate scars, while Ref.~\cite{cenedese2024} has generated scar states by manually engineering conserved quantities.
Despite these advances, many other non-thermal states in these systems remain unexplored, highlighting the need for further investigation into hidden eigenstates that could exhibit atypical properties.

\begin{figure}[tb]
\begin{center}
\includegraphics[clip = true, width =\columnwidth]{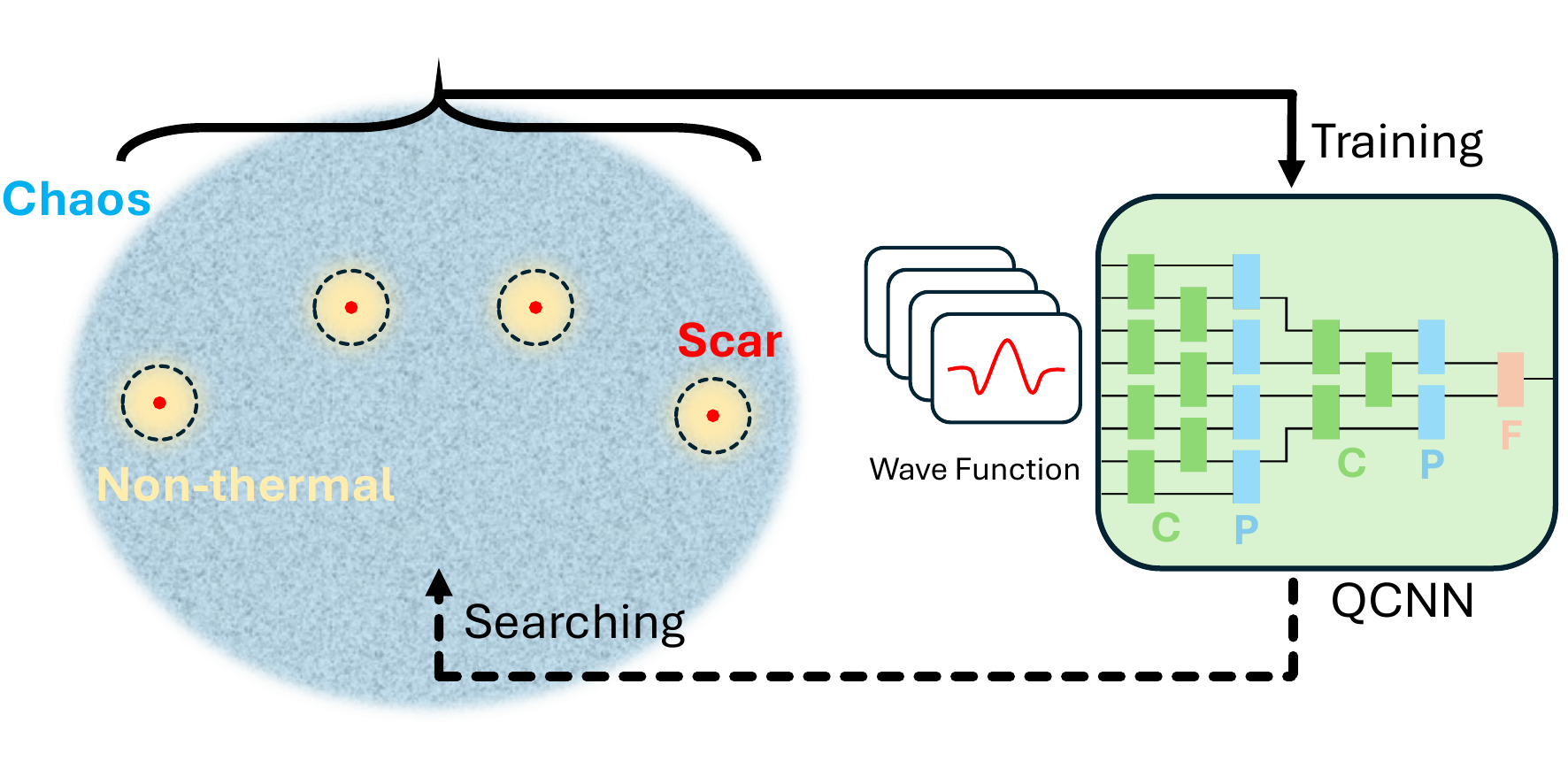}
\caption{\label{fig:setup}  Conceptual plot of the non-thermal states classified by QCNN. The wave functions inside and outside the scar subspace are treated as two classes that are input to the QCNN as training data. The QCNN consists of convolution layers(C), pooling layers(P) and a fully connected layer(F). Then QCNN is able to identify a larger subspace (dashed circles) of special states that are potential candidates for approximate scars.}
\end{center}
\end{figure}

We investigate non-thermal states in chaotic systems by resorting to QCNNs. As depicted in Fig.~\ref{fig:setup}, our approach involves training the QCNN on a subset of states within the known subspace of QMBSs.
In order to explore the potential of QCNNs in identifying non-thermal states, we investigate three models known to harbor QMBS: xorX model~\cite{Iadecola2020}, PXP model~\cite{Turner2018,Lin2019} and far-coupling Ising Su-Schrieffer-Heeger(SSH) model~\cite{ZhangPengfei2023}. In simulations, the QCNN achieves over 99\% single-shot measurement accuracy in identifying all known QMBS states. Moreover, it uncovers additional non-thermal states with scar-like properties that extend beyond the established QMBS families.
In the xorX model, some of these non-thermal states can be approximately described by spin-wave modes associated with specific quasiparticles. To capture key features of these many-body eigenstates, we develop effective tight-binding Hamiltonians within the quasiparticle subspace. We further validate the QCNN's performance experimentally on IBM quantum devices. By employing error mitigation techniques, our QCNN achieves a single-shot measurement accuracy of around $63\%$.

\section{RESULTS}
\subsection{Models}
In this work, we consider three different systems harboring QMBSs, as detailed in this section. Our primary focus is the xorX model.
In this model, single spin flip (X) occurs when its nearest neighbors satisfy the exclusive or (xor) condition. The xorX model stands out because it allows for the analytical solution of a family of exact scar states, providing a clear testbed for exploring non-thermal phenomena, though it remains an open question whether other types of scar states exist.
The xorX model under open boundary condition (OBC) is described by the Hamiltonian~\cite{Iadecola2020, cenedese2024}
\be
  H_{\rm xorX}= \lambda\sum_{i=2}^{n-1}\left( \sigma_i^x-\sigma_{i-1}^z \sigma_i^x \sigma_{i+1}^z \right)+\Delta\sum_{i=1}^{n}\sigma_i^z+J\sum_{i=1}^{n-1}\sigma_i^z\sigma_{i+1}^z ~,
\label{eq:xorX}
\ee
where $\sigma_i^{x}, \sigma_i^{y}, \sigma_i^{z}$ are the Pauli-$X,Y,Z$ matrices for the $i$-th qubit and $n$ is the total number of qubits.
In xorX model, the boundary qubits ($i=1,n$) are frozen since $\left[ H,\sigma_{1}^z \right]=\left[ H,\sigma_{n}^z \right]=0$. We focus on the subspace of $\left\langle\sigma_{1}^z\right\rangle=\left\langle\sigma_{n}^z\right\rangle=-1$. A family of exact scar states in the xorX model can be identified as~\cite{Iadecola2020}
\begin{eqnarray}
|\mathcal{S}_m\rangle &=& \frac{1}{m!\sqrt{\mathcal{N}_m}}\left( Q^\dagger\right)^m|0\rangle^{\otimes n} ~,
\label{eq:ExaScar}
\end{eqnarray}
where $\mathcal{N}_m$ is the normalization factor and the operator
\begin{eqnarray}
 Q^\dagger &=& \sum_{i=2}^{n-1}(-1)^iP_{i-1}^0\sigma_i^+P_{i+1}^0 ~,
\end{eqnarray}
with projectors $P_{i}^0=|0\rangle_i\langle0|$ and $P_{i}^1=|1\rangle_i\langle1|$.
The domain wall number in the xorX model is conserved as $\left[ H,\sum_i\sigma_i^z\sigma_{i+1}^z \right]=0$.

The second model we consider in this work is the PXP model, which is derived from the Rydberg atom system in the Rydberg blockade regime \cite{Turner2018}.
The Hamiltonian of the PXP model is~\cite{Turner2018,Lin2019}
\begin{eqnarray}
H_{\rm PXP}=\frac{\Omega}{2}\sum_{i=2}^{n-1}P_{i-1}^0\sigma_i^xP_{i+1}^0 ~,
\end{eqnarray}
where $\Omega$ represents the overall energy scale. Notably, the known scar states exhibit a large overlap with the anti-ferromagnetic (Néel) state $|Z_2\rangle$, commonly referred to as $Z_2$ tower states, which accounts for the persistent oscillations observed in experiments~\cite{bernien2017probing}. Although a few scar states can be analytically solved using matrix product states~\cite{Lin2019}, the nature of other scar states remains an active area of research \cite{ivanov2024volumepxp,WangJiaWei2024,Deger2024,HaRu2024}.

The far-coupling Ising SSH model is realized on the platform of superconducting circuit~\cite{ZhangPengfei2023}. The serpentine routing makes it flexible to tune the coupling between different qubits. The Hamiltonian is
\begin{eqnarray}
H_{\rm fc}&=&\sum_{i=1}^{\left\lfloor\frac{n-1}{2}\right\rfloor}\left(J_{\rm e}\sigma_{2i-1}^+\sigma_{2i}^-+J_{\rm o}\sigma_{2i}^+\sigma_{2i+1}^-\right)  \nonumber\\
&&+J_{\rm nn}\sum_{i=1}^{n-3}\sigma_i^+\sigma_{i+3}^-+{\rm h.c.} ~.
\end{eqnarray}
where $\sigma^+=|1\rangle\langle0|$ and $\sigma^-=|0\rangle\langle1|$ represent the raising and lowering operators, respectively. $J_{\rm e}$ and $J_{\rm o}$ denote the coupling strengths at even and odd positions. $J_{\rm nn}$ is the next-next-nearest-neighbor coupling strength which breaks integrability.
Both numerical simulations and experimental data provide evidence for the existence of scar states in this model, which exhibit a significant overlap with the reference state $|Z_{1001}\rangle=|1001\rangle^{\otimes n/4}$.

\begin{figure}[tb]
\begin{center}
\includegraphics[clip = true, width =\columnwidth]{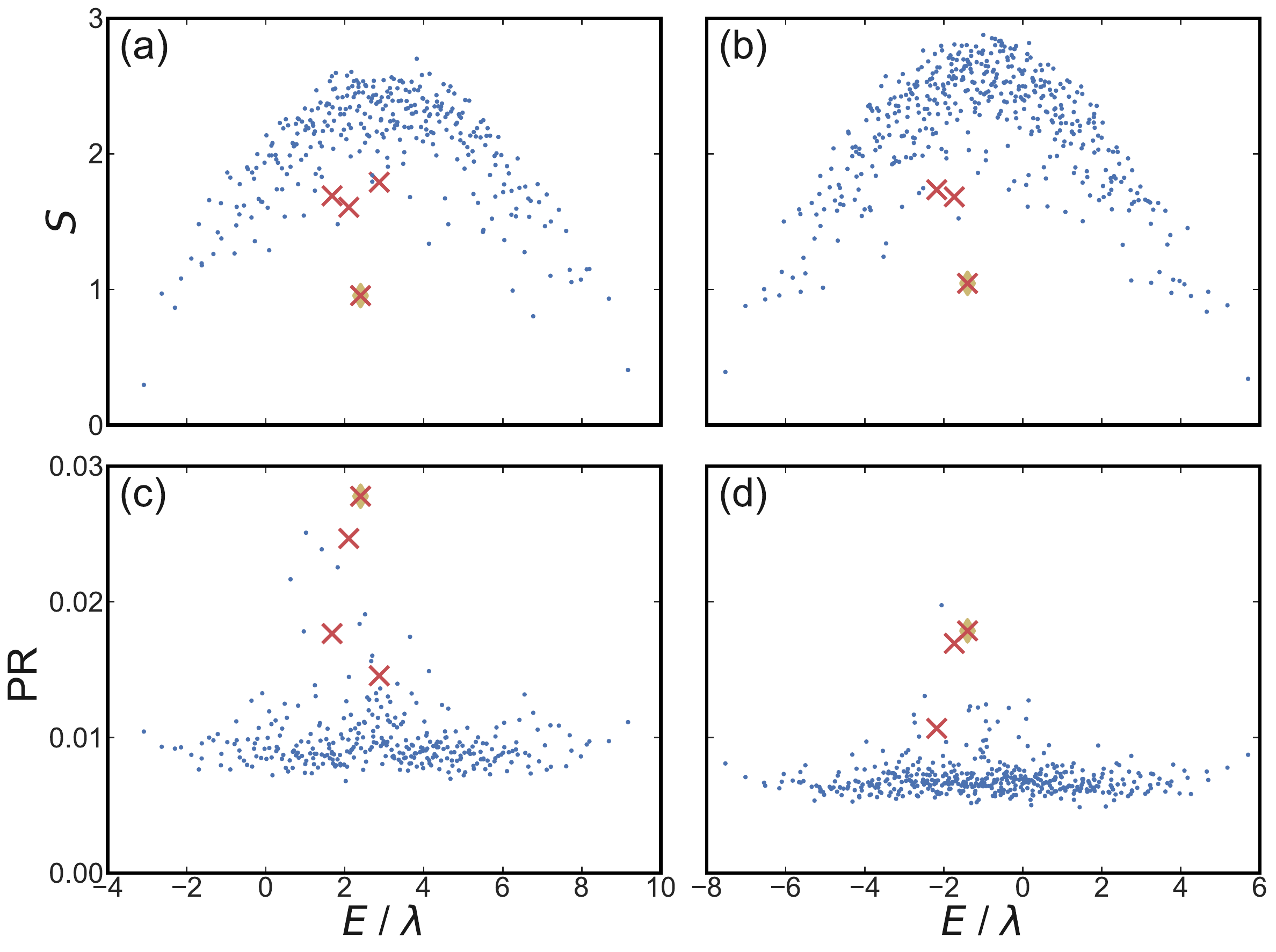}
\caption{\label{fig:S_IPR}  The entanglement entropy and participation ratio of the eigenstates of xorX model within different domain-wall number sectors. (a)(c) $n_{\rm dw}=2$, (b)(d) $n_{\rm dw}=3$. Other parameters are $\lambda=J=10\Delta$. The red crosses are the eigenstates marked by the QCNN. The prisms are the exact scar states. The number of spins is 12 while two of them at the boundaries are fixed.}
\end{center}
\end{figure}

\subsection{Non-thermal states}
We begin our study with the xorX model in Eq.~\eqref{eq:xorX}, where there is a family of well-defined exact scar states. Since the exact scar states Eq.~\eqref{eq:ExaScar} are independent of the parameters in the Hamiltonian Eq.~\eqref{eq:xorX}, the trained QCNN is also parameter independent. Before experimental implementation that will be presented in section \emph{Experimental demonstration on quantum device}, we first perform numerical simulations on classical computers.
After training, the quantum circuit classifies the eigenstates into two types. Interestingly, while the total loss decays during the training, the final converged loss remains near 0.14. Moreover, it successfully recognizes all the exact scar states with an error probability of single-shot measurement less than 1\%. This means that the QCNN definitely recognizes all the exact scar state and is expected to do so with sufficient measurement in experiment \cite{Harry2022}. The disparity between the large loss function and the high accuracy in recognizing scar states indicates that the QCNN also classifies some additional states, beyond the known exact scars, as ``scar" states. As we detail in section \emph{Spin-wave approximation for the marked states in xorX model}, we identify a substantial portion of these states as non-thermal states, which bear a significant resemblance to the exact scar states.

\begin{figure}[tb]
\begin{center}
\includegraphics[clip = true, width =\columnwidth]{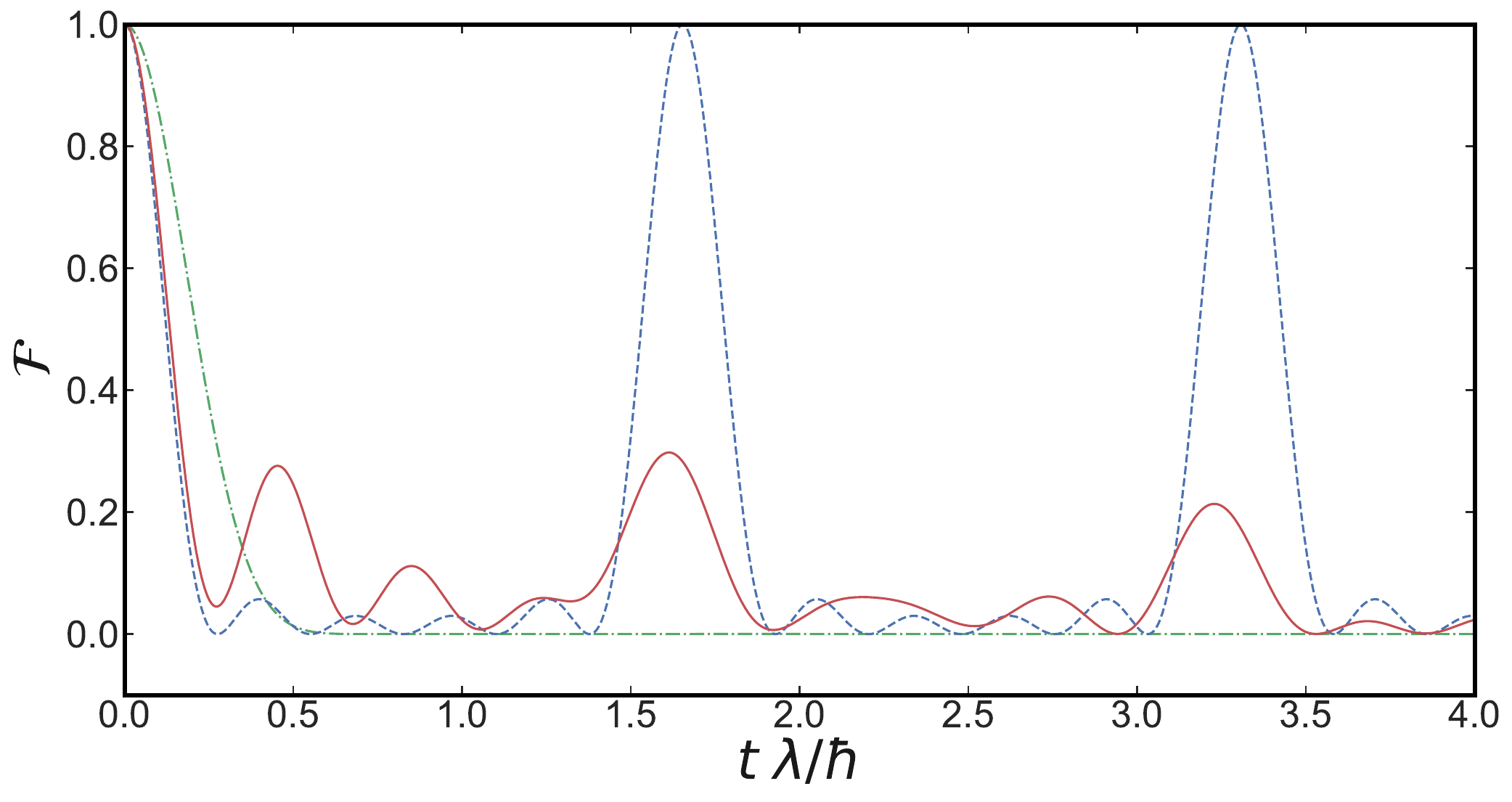}
\caption{\label{fig:Revival} Revival behavior under Hamiltonian evolution. The plot shows fidelity as a function of time, with the initial state being an equal-amplitude superposition of exact scar states (blue dashed line), additional states identified by QCNN (red solid line), and non-marked states (green dash-dotted line). The parameters are the same as in Figure 2. }
\end{center}
\end{figure}

\begin{figure}[tb]
\begin{center}
\includegraphics[clip = true, width =\columnwidth]{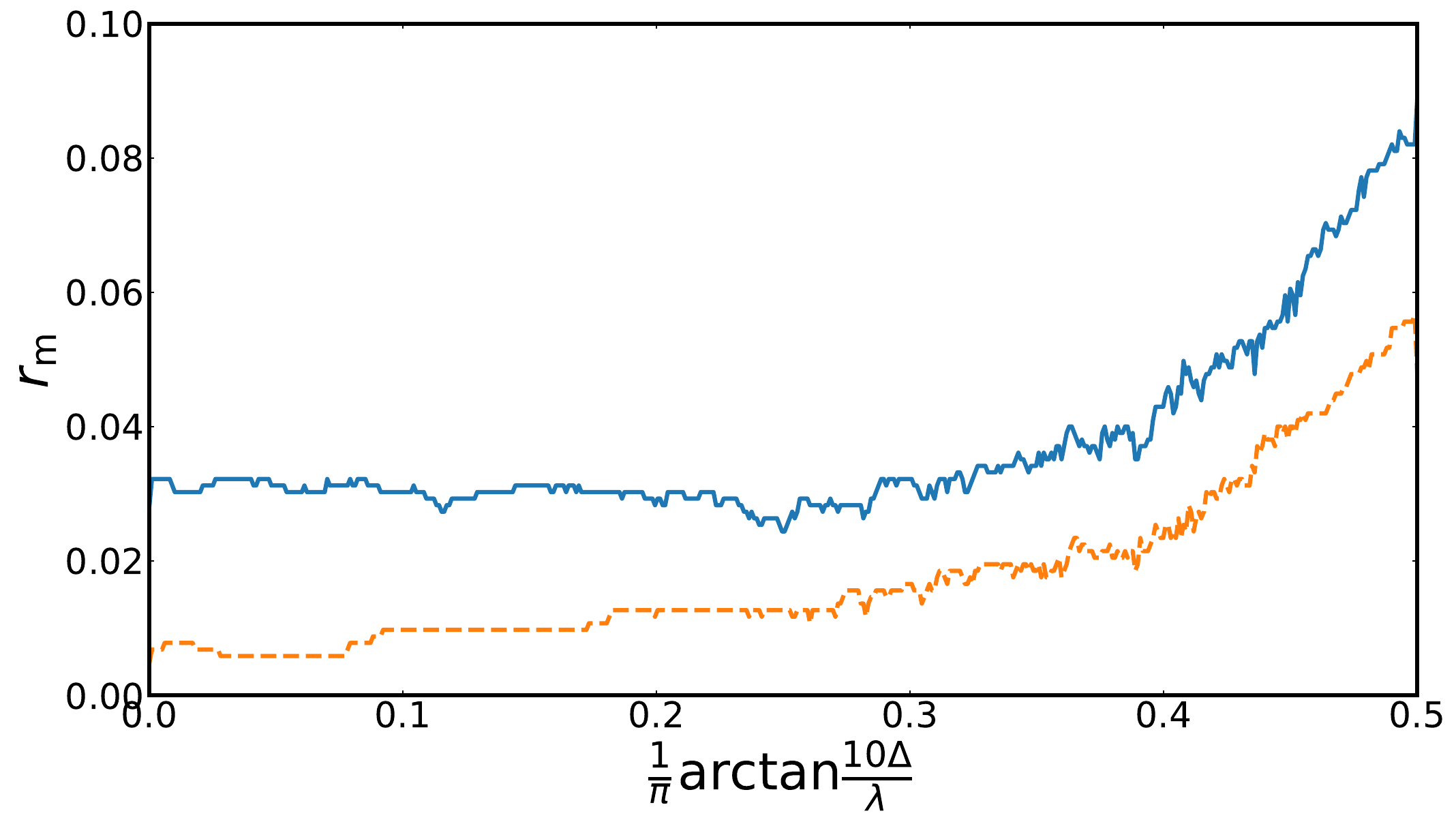}
\caption{\label{fig:MarNum}  The proportion of marked eigenstates identified by the QCNN to the total eigenstates. The results are shown for architectures with 2 (blue solid line) and 3 (orange dashed line) convolutional layers before the pooling layers.}
\end{center}
\end{figure}

The additional non-thermal states have similar energy as the exact scar states, as shown in Fig. \ref{fig:S_IPR}. These states are situated in the middle of the energy spectrum, distinguishing them from the low-energy integrable modes. Their half-chain entanglement entropy is lower than that of the bulk chaotic states, as shown in Fig.~\ref{fig:S_IPR}(a)(b), indicating potential deviation from the volume-law entanglement entropy. The QCNN can compensate for deficiencies that the entanglement entropy may fail to distinguish states \cite{Gu2024}. Additionally, the participation ratio (PR), defined as $\sum_i\left|\langle\psi|i\rangle\right|^4$ for a given state $\ket{\psi}$ in the computational basis $\{\ket{i}\}$, is significantly higher than that of the majority of chaotic states, as shown in Fig.~\ref{fig:S_IPR}(c)(d). They are thus constrained within a smaller Hilbert space compared to chaotic states \cite{Mondragon2021}.

In addition to the static metrics presented above, the existence of scars is often demonstrated through revivals of fidelity in quench dynamics.
Here we evolve an initial state under the xorX Hamiltonian in Eq.~\eqref{eq:xorX}. In Fig.~\ref{fig:Revival}, we plot the fidelity of the initial state $\mathscr{F}=\left| \langle \psi_0|\psi_t\rangle\right|^2$ as a function of time, for three different choices of initial states.
As expected, an equal superposition of all known exact scar states shows perfect revivals, as illustrated by the blue dashed curve. The superposition of the additional non-thermal states identified by the enhanced QCNN also exhibits revivals, though with a decaying amplitude and not strictly periodic oscillations, as indicated by the red solid curve. 
In contrast, the superposition of non-marked states does not exhibit any revival (green dashed), as is expected for generic quantum chaotic dynamics. Additional cases of fidelity oscillations are detailed in Appendix~\ref{App:QCNN}. 
Such revivals in the fidelity of initial states highlight a clear distinction between states marked by the QCNN and generic chaotic states, providing further evidence for the non-thermal characteristics of the former.

While the exact scar states of Eq.~\eqref{eq:ExaScar} in the xorX model are parameter independent, we expect that the fraction of these additional non-thermal states can be tuned by varying certain parameters of Hamiltonian~(\ref{eq:xorX}). For example, upon increasing $\Delta$, pairs of domain walls become more and more confined, which leads to slow thermalization and non-ergodic dynamics~\cite{PhysRevLett.122.150601}.
In Fig.~\ref{fig:MarNum}, we confirm that the ratio of non-thermal states identified by the enhanced QCNN increases with $\Delta$. This further demonstrates the enhanced QCNN's ability to discern atypical states from the eigenspectrum. On the other hand, when the circuit has more parameters, the criterion becomes stricter. Only states that are close enough to the exact scar states will be marked \cite{Wang2021D}. As a result, the ratio of non-thermal states is smaller, as indicated by the orange dashed line being lower than the blue solid line in Fig.~\ref{fig:MarNum}.

\subsection{Spin-wave approximation for the marked states in xorX model}
\label{sec:theory}
The numerical results presented in the previous section suggest a more detailed study on the nature of the additional non-thermal states found by QCNN in the xorX model in Eq.~\eqref{eq:xorX}. The PR indicates that these states are predominantly localized within a small subregion of the full Hilbert space. In this section, we demonstrate that some of these non-thermal states can be understood in terms of quasiparticles, specifically magnon bound states. We will construct effective tight-binding Hamiltonians that approximately describe the spin-wave modes of these quasiparticles, reproducing key features of the exact many-body eigenstates.

\subsubsection{Integrable states}

\begin{figure}[tb]
\begin{center}
\includegraphics[clip = true, width =\columnwidth]{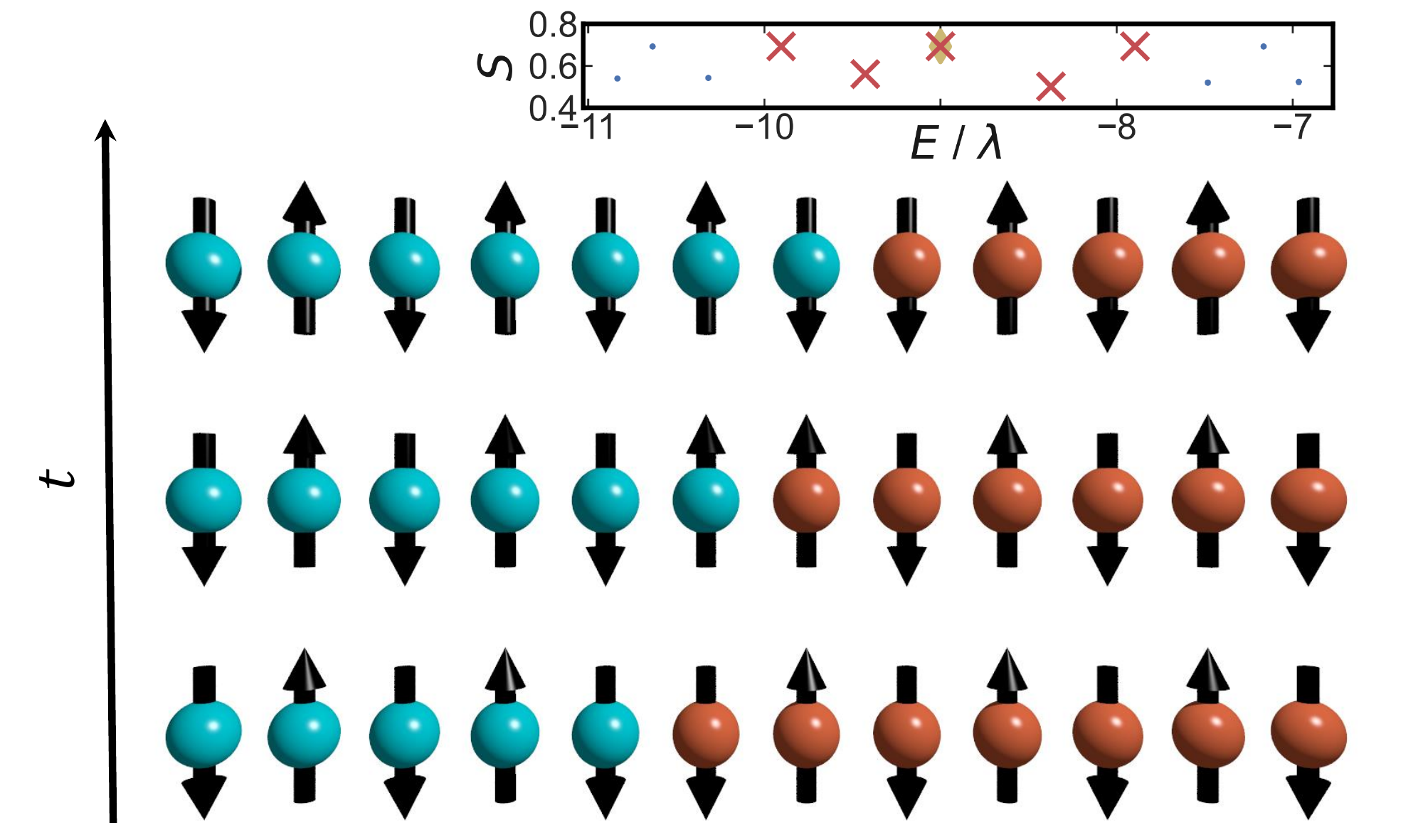}
\caption{\label{fig:AFSpinWave}  Domain wall dynamics between two antiferromagnetically ordered domains, with the inset highlighting the eigenstates within this subspace. These eigenstates are characterized by distinct eigen-energies with similar entanglement entropy.}
\end{center}
\end{figure}
We begin by considering the simplest scenario. The sequence of exact scar states $|\mathcal{S}_m\rangle$ in Eq.~(\ref{eq:ExaScar}) satisfies $m\leq n/2$ for a system of $n$ spins. In particular, the state $|\mathcal{S}_{\lfloor n/2\rfloor}\rangle$ is an anti-ferromagnetic state residing close to the edge of the energy spectrum. For $n$ even, the configuration consistent with the boundary conditions features two domains with different anti-ferromagnetic orders, separated by a single domain wall, as shown in Fig.~\ref{fig:AFSpinWave}. The Hamiltonian of Eq.~(\ref{eq:xorX}) acting on this configuration generates a hopping term for the single domain wall and a staggered on-site potential that depends on the sublattice where the domain wall resides.
This leads to the following effective single-particle Hamiltonian within the subspace defined by a single domain wall separating two anti-ferromagnetic domains:
\begin{eqnarray}
    H=\sum_{i=1}^{n-1}(-1)^i \Delta d_i^\dagger d_i+\lambda d_{i+1}^\dagger d_i+\lambda d_i^\dagger d_{i+1} ~.
\end{eqnarray}
where $i$ is the position of the domain wall.
In this subspace, the dynamics is fully integrable.

The above effective Hamiltonian can be readily diagonalized. Under periodic boundary conditions(PBC), the Hamiltonian in momentum space takes the form:
\begin{eqnarray}
H &=& \left(\begin{matrix}
        \Delta & \lambda+\lambda e^{ik} \\
        \lambda+\lambda e^{-ik} & -\Delta
      \end{matrix}\right) ~,
\label{eq:H1}
\end{eqnarray}
where we have set the lattice spacing to unity.
The eigenenergies are $E_k=\pm \sqrt{\Delta^2+4\lambda^2\cos ^2(k/2)}$, with corresponding eigenstates $\phi_k$. Under OBC, the eigenstates approximate standing waves, expressed as superpositions of $\phi_k$ and $\phi_{-k}$. Specifically, these superpositions take the form of $\left(\ket{\phi_k}+\ket{\phi_{-k}} \right)/\sqrt{2}$ and $\left(\ket{\phi_k}-\ket{\phi_{-k}} \right)/\sqrt{2}$.
These states exhibit low entanglement entropy, characteristic of integrable systems, as shown in the inset of Fig. \ref{fig:AFSpinWave}.
 The QCNN successfully identifies states within this integrable subspace, marking those with energies similar to the exact scar states.


\begin{figure}[tb]
\begin{center}
\includegraphics[clip = true, width =\columnwidth]{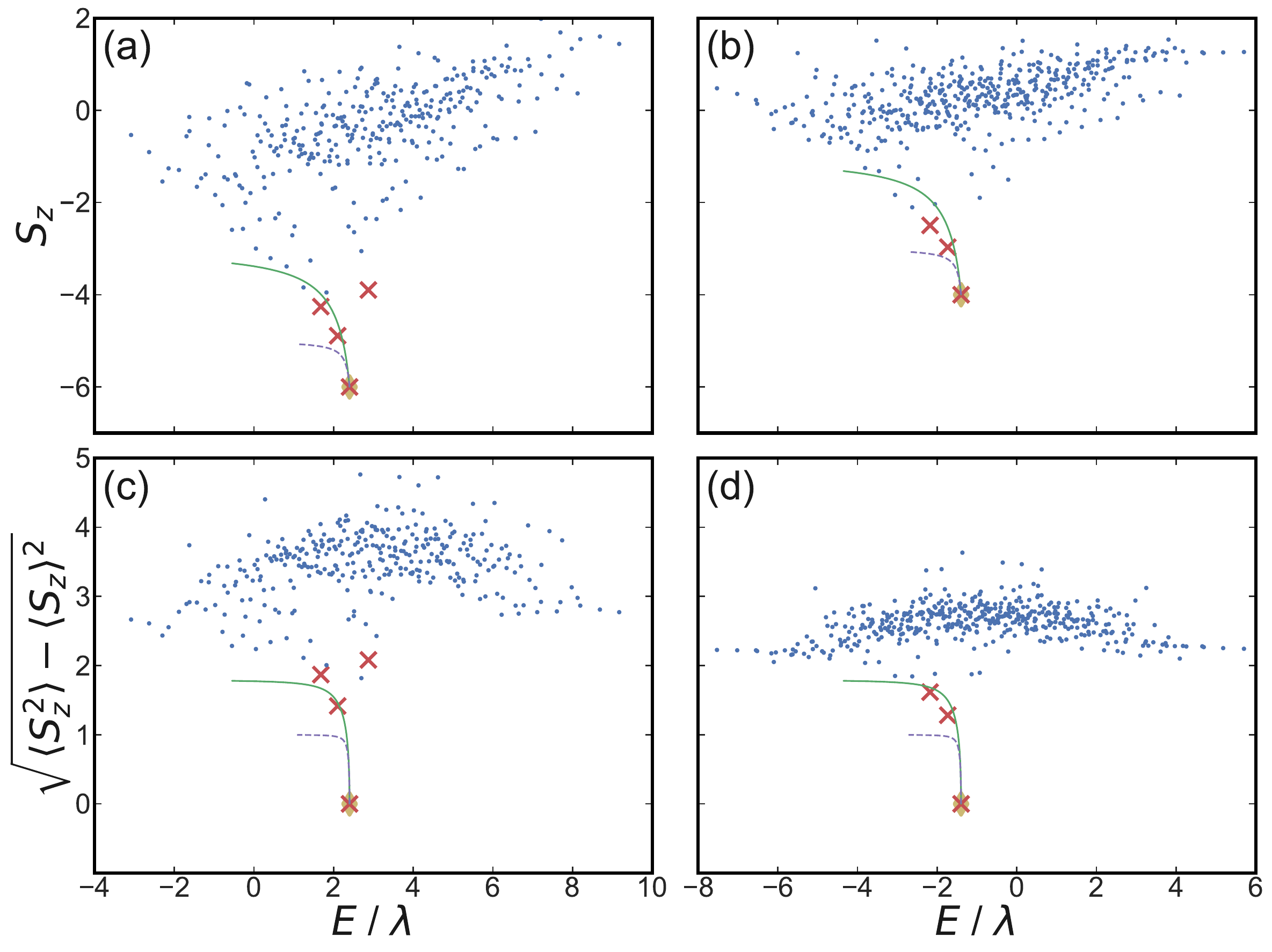}
\caption{\label{fig:Integ01}  The mean and variance of the total $S_z$ for each eigenstate within a certain $n_{\rm dw}$ sector: (a)\&(c) $n_{\rm dw}=2$; (b)\&(d) $n_{\rm dw}=3$. The red crosses indicate the eigenstates identified by the QCNN. The green solid curves represent analytical results from the ferromagnetic magnon bound state approximation, while the purple dashed curves correspond to those obtained from the anti-ferromagnetic magnon bound state approximation. Other parameters are $\lambda=J=10\Delta$. }
\end{center}
\end{figure}


\subsubsection{Approximate quasiparticle states}

\begin{figure}[tb]
\begin{center}
\includegraphics[clip = true, width =\columnwidth]{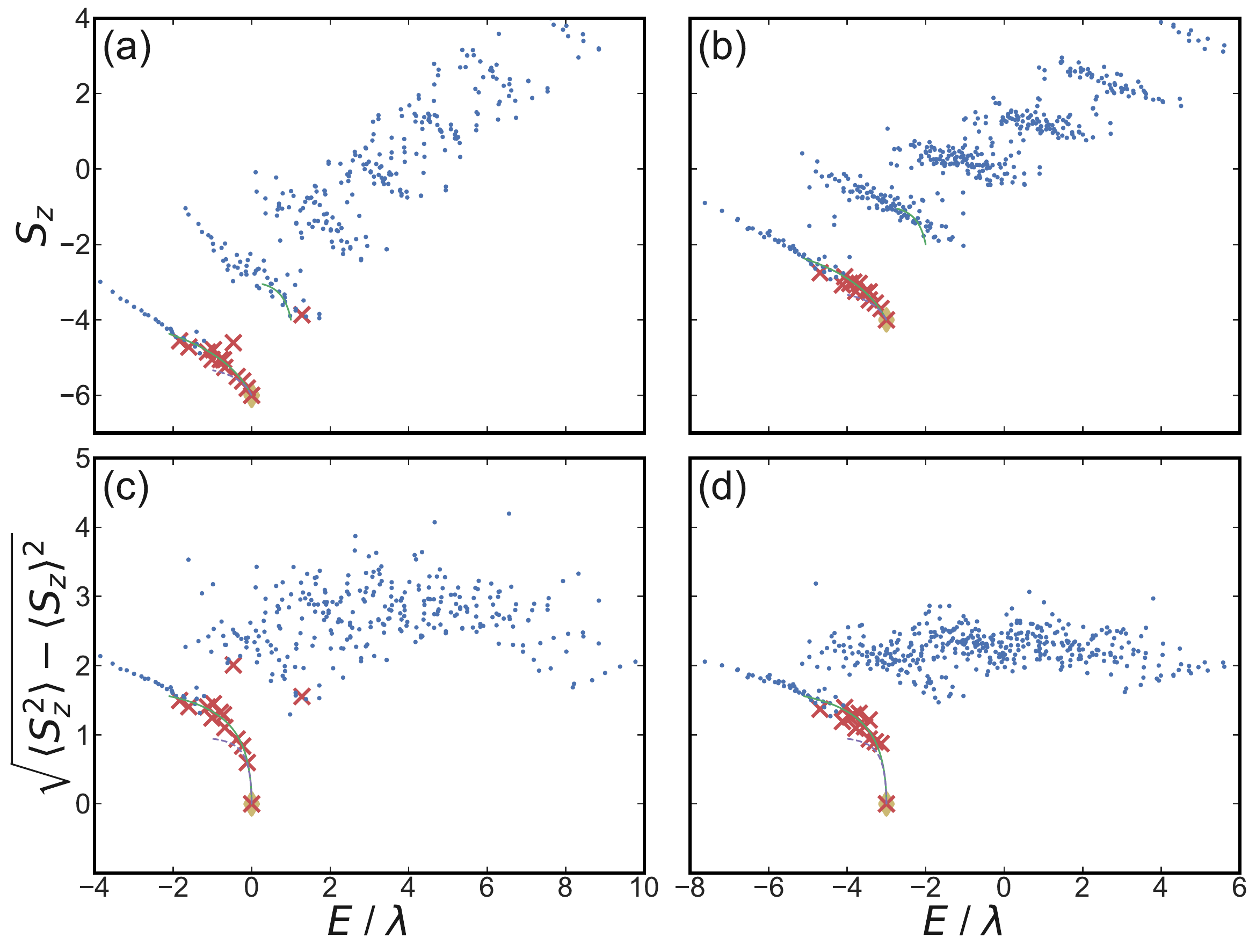}
\caption{\label{fig:Integ05}  The mean and variance of the total $S^z$ for each eigenstate within a certain $n_{\rm dw}$ sector: (a)\&(c) $n_{\rm dw}=2$; (b)\&(d) $n_{\rm dw}=3$. The red crosses indicate the eigenstates identified by the QCNN. The green solid curves represent analytical results from the ferromagnetic magnon bound state approximation, while the purple dashed curves correspond to those obtained from the anti-ferromagnetic magnon bound state approximation. Other parameters are $\lambda=J=2\Delta$.}
\end{center}
\end{figure}

To understand the nature of the states marked as `scar like' by QCNN, we first calculate the mean and variance of their total $z$-magnetization $S_z=\sum_{i=2}^{n-1}\sigma_i^z$, and compare  these values with those of typical thermal eigenstates. Fig.~\ref{fig:Integ01} presents the results for two different domain wall number $n_{\rm dw}$ sectors. The states marked by QCNN (red crosses) exhibit both a lower average magnetization and a smaller variance compared to typical eigenstates. In the parameter regime with larger $\Delta$, where spin flipping becomes more difficult, the system exhibits increased integrability and better conservation of $S_z$. Consequently, more eigenstates are marked by QCNN, as shown in Fig. \ref{fig:Integ05}.

\begin{figure}[tb]
\begin{center}
\includegraphics[clip = true, width =\columnwidth]{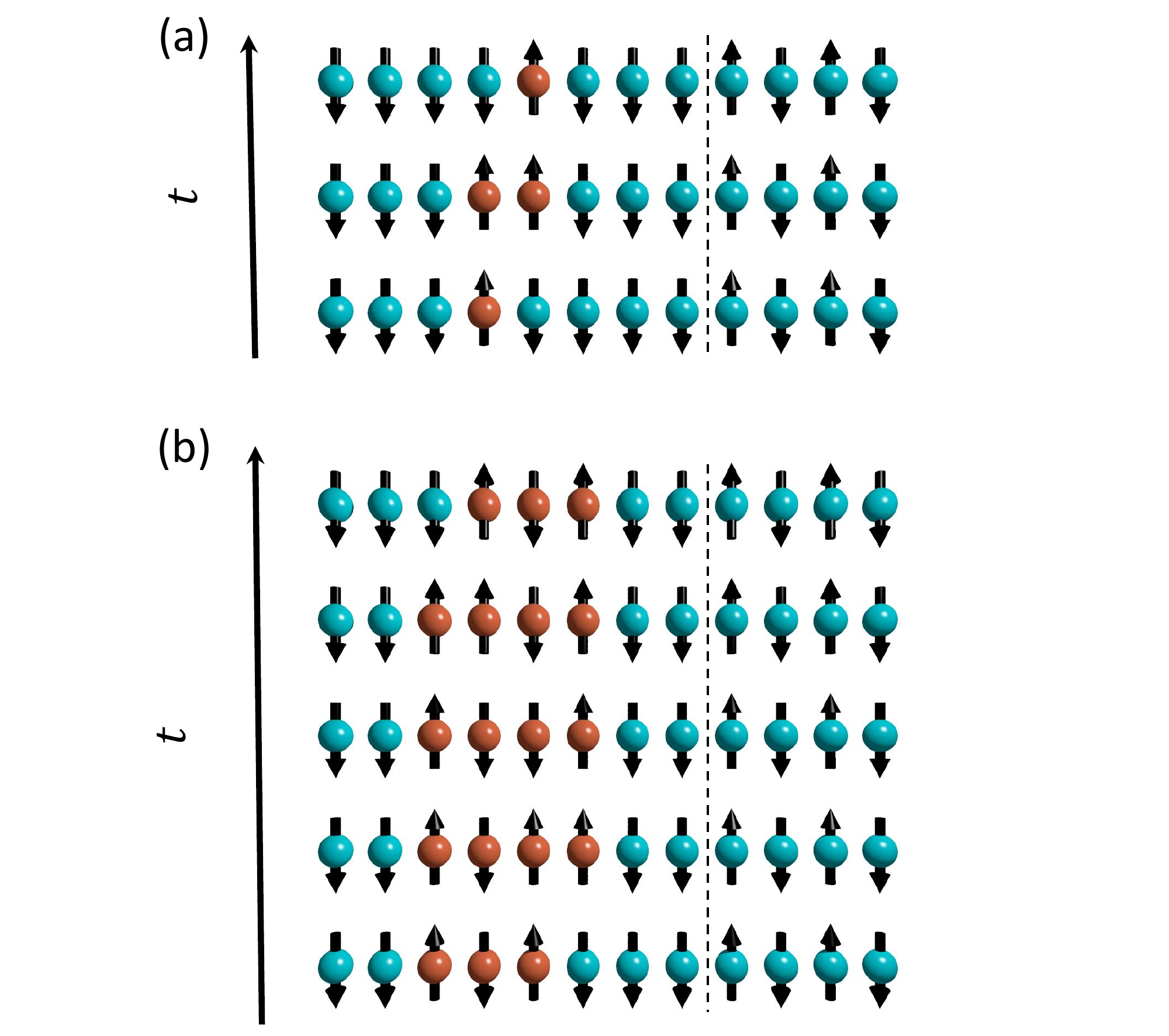}
\caption{\label{fig:moving}  Illustration of two types of quasiparticles moving via intermediate processes. On the left side of the dashed line, blue spins form a ferromagnetic background for the propagation of the bound state. On the right side of the dashed line, blue spins are trapped in an antiferromagnetic configuration.
  (a) The motion of a ferromagnetic bound state (red spins). (b) The motion of an anti-ferromagnetic bound state (red spins). }
\end{center}
\end{figure}

This strongly suggests that these states exhibit a special structure: they can be interpreted as quasiparticles moving within an almost ferromagnetically ordered background (with a negative net magnetization). However, there are two  critical differences compared to the quasiparticles in the tower of exact scar states of Eq.~(\ref{eq:ExaScar}) and the fully integrable states discussed in the previous section. First, the quasiparticle picture is only approximate. While the exact eigenstates predominantly reside within the quasiparticle subspace, they also have non-negligible components in other configurations within the Hilbert space (see Fig.~\ref{fig:P010}). Second, the quasiparticles in this system are generally more intricate than single magnons or domain walls, often involving longer strings or more complex structures. Moreover, the motion of these quasiparticles typically includes intermediate  stages where the size of the quasiparticles first grows and then shrinks (see Fig.~\ref{fig:moving}).
In the following, we construct effective Hamiltonians to describe the approximate spin-wave modes of these quasiparticles and demonstrate that they capture similar key features of the exact many-body eigenstates.

\subsubsection{Ferromagnetic magnon bound state}

\begin{figure}[tb]
\begin{center}
\includegraphics[clip = true, width =\columnwidth]{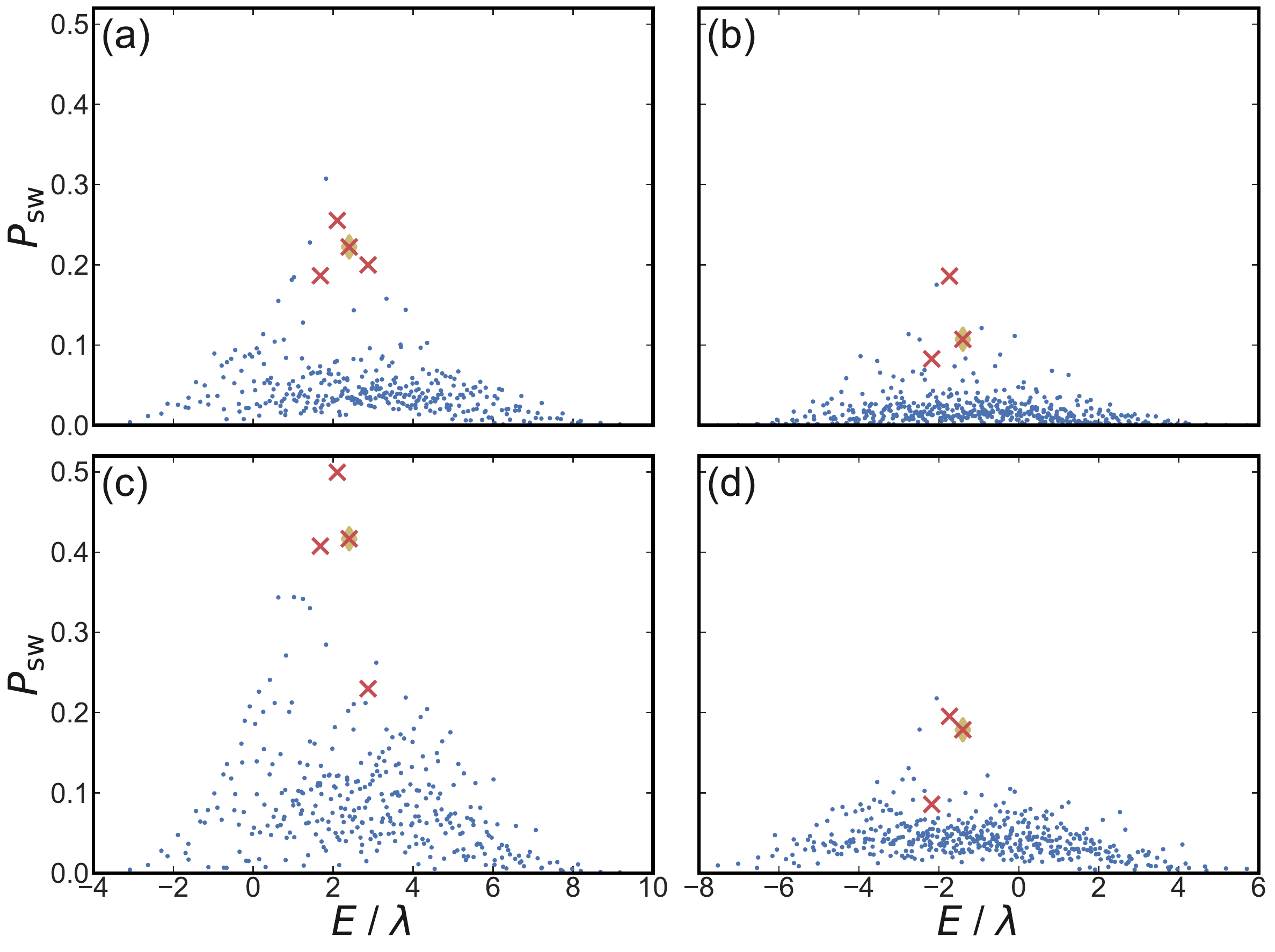}
\caption{\label{fig:P010}  The spin-wave component in each eigenstate. The weights of the ferromagnetic magnon bound state are shown in (a) and (b), while the weights of the anti-ferromagnetic magnon bound state are shown in (c) and (d). The special states marked by QCNN have anomalously large weight on the subspace of a particular type of quasiparticle, compared to typical eigenstates.
Other parameters are $\lambda=J=10\Delta$, (a)(c)$n_{\rm dw}=2$, (b)(d)$n_{\rm dw}=3$.}
\end{center}
\end{figure}

We begin by analyzing the motion of a single magnon in a background of down spins, as depicted in Fig.~\ref{fig:moving}(a). In Fig.~\ref{fig:P010}(a)\&(b), we present the total weight of each eigenstate within the single-magnon configuration subspace. The data reveal that certain marked states exhibit significantly larger weights in this subspace. The motion of this single magnon will necessarily involve intermediate configurations where the magnon first grows into longer strings and then shrinks. For instance, consider the following intermediate configurations (totaling four configurations): $\{\cdots00111100\cdots$, $\cdots0011100\cdots$, $\cdots001100\cdots$, and $\cdots00100\cdots\}$. The effective Hamiltonian within this subspace has the following form in momentum space (assuming PBC):
\begin{eqnarray}
  H &=& \left(\begin{matrix}
                3\Delta & \lambda+\lambda e^{ik} & 0 & 0 \\
                \lambda+\lambda e^{-ik} & \Delta & \lambda+\lambda e^{ik} & 0 \\
                0 & \lambda+\lambda e^{-ik} & -\Delta & \lambda+\lambda e^{ik} \\
                0 & 0 & \lambda+\lambda e^{-ik} & -3\Delta
              \end{matrix}\right).
\label{eq:fe}
\end{eqnarray}

The analytical solution of the ground state energy is $E_k=-\Delta\sqrt{10+3u^2+\sqrt{64+48u^2+5u^4}}/\sqrt{2}$, where $u=\left|\lambda+\lambda e^{ik}\right|/\Delta$. Under OBC, the system approximately forms standing waves as a superposition of states with momenta $\pm k$. By varying $k$, we calculate $E_k$ and the corresponding $S_z$. The relationship between these quantities is illustrated by the green solid curves in Figs.~\ref{fig:Integ01} and \ref{fig:Integ05}.
Most of the marked states closely align with these curves, suggesting that they can indeed be interpreted as spin-wave modes. Furthermore, as $\Delta$ increases, the agreement between the marked states and the analytical approximation improves, as shown in Fig. \ref{fig:Integ05}.

\subsubsection{Anti-ferromagnetic magnon bound state}

We identify another component of special states, recognized by QCNN, which can be understood as quasiparticles of a short anti-ferromagnetic string, as depicted in Fig.~\ref{fig:moving}(b). In Fig.~\ref{fig:P010}(c)\&(d), we plot the total weight of each eigenstate in the subspace of the shortest  anti-ferromagnetic string. The results confirm that some marked states exhibit unusually large weights in these configurations compared to typical eigenstates.  Restricting to the subspace of the four configurations shown in Fig.~\ref{fig:moving}(b), we can similarly write down an effective Hamiltonian:
\begin{eqnarray}
  H &=& \left(\begin{matrix}
                \Delta & \lambda & 0 & \lambda e^{ik} \\
                \lambda & -\Delta & \lambda & 0 \\
                0 & \lambda & \Delta & \lambda \\
                \lambda e^{-ik} & 0 & \lambda & -\Delta
              \end{matrix}\right).
\label{eq:AF}
\end{eqnarray}
The energy of the ground state and the first excited state are $E_k=-\sqrt{\Delta^2+2\lambda^2(1\pm\cos (k/2))}$. Since the energy of the first excited state is closer to the exact scar state, we present it with purple dashed curves in Figs.~\ref{fig:Integ01} and \ref{fig:Integ05}. The deviation of the antiferromagnetic magnon bound states from the exact eigenstates is greater than that of the ferromagnetic magnon bound states, leading to a smaller component contribution, as shown in Fig. \ref{fig:P010}.

It is worth emphasizing that there are also some integrable local modes found in this model in the low-energy regime~\cite{Birnkammer2022}. They can be approximated by oscillators in a linear potential, which give rise to spatially localized modes. Such states with localized modes have a rather distinct nature compared to the tower of states in Eq.~(\ref{eq:ExaScar}) that we use as training set. Indeed, these trivial states are not marked by the QCNN. The non-thermal states discussed in this section, in contrast, are situated in the middle of the energy spectrum.

\subsection{Experimental demonstration on quantum device}
\label{sec:experiment}
We demonstrate the performance of our QCNN on IBM's quantum hardware, with the training process carried out classically via noise-free simulations. We then prepare the exact scar state $|\mathcal{S}_1\rangle$ using a shallow circuit, which is fed into the trained QCNN on the quantum device to evaluate its performance. To combat noise, we introduce a shallow general layer as a preprocessing step to enhance hardware efficiency. Furthermore, the learning circuit is optimized by reducing the number of two-qubit gates, improving overall implementation.

\begin{figure}[tb]
\begin{center}
\includegraphics[clip = true, width =\columnwidth]{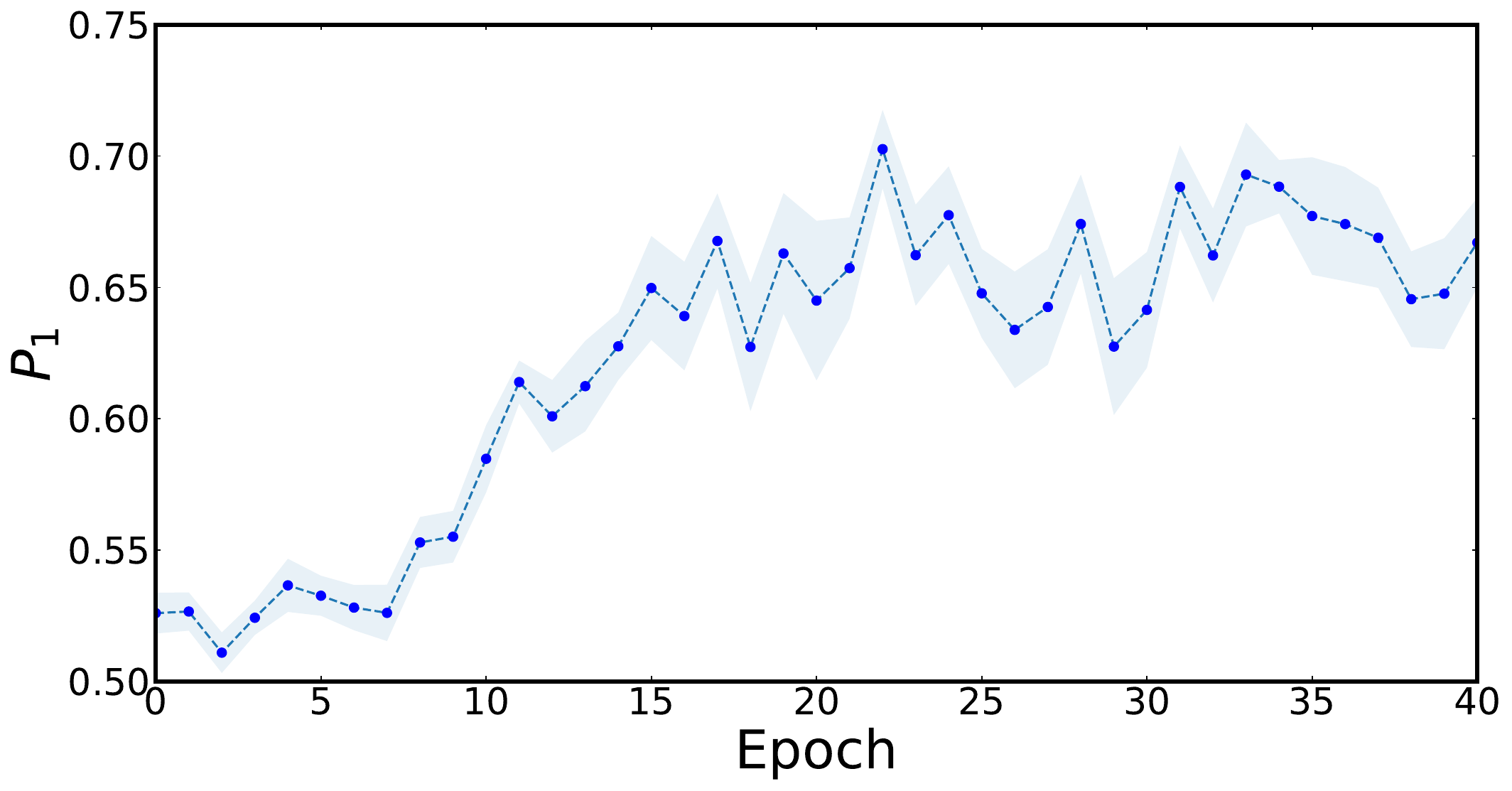}
\caption{\label{fig:ExperimentLoss}  Success rate of the QCNN during training on a quantum device. The experiment was conducted in 12 groups, each spaced hours apart to assess drift error. Each group involved $10^4$ measurement repeats to determine the success probability. Error bars indicated by the shaded region represent the standard error across different groups. }
\end{center}
\end{figure}

The circuit used to prepare the $|\mathcal{S}_1\rangle$ state is depicted in Appendix~\ref{App:Exp}.
The trained QCNN successfully identifies this state, achieving a success rate of over 99\% in a noiseless classical simulation.
The success rate observed on quantum hardware is presented in Fig. \ref{fig:ExperimentLoss}, which improves as the number of iterations in the learning process increases. However, due to the inherent noise in real-world quantum devices, the overall success rate is lower than that achieved in noiseless simulations.

\begin{figure}[tb]
\begin{center}
\includegraphics[clip = true, width =\columnwidth]{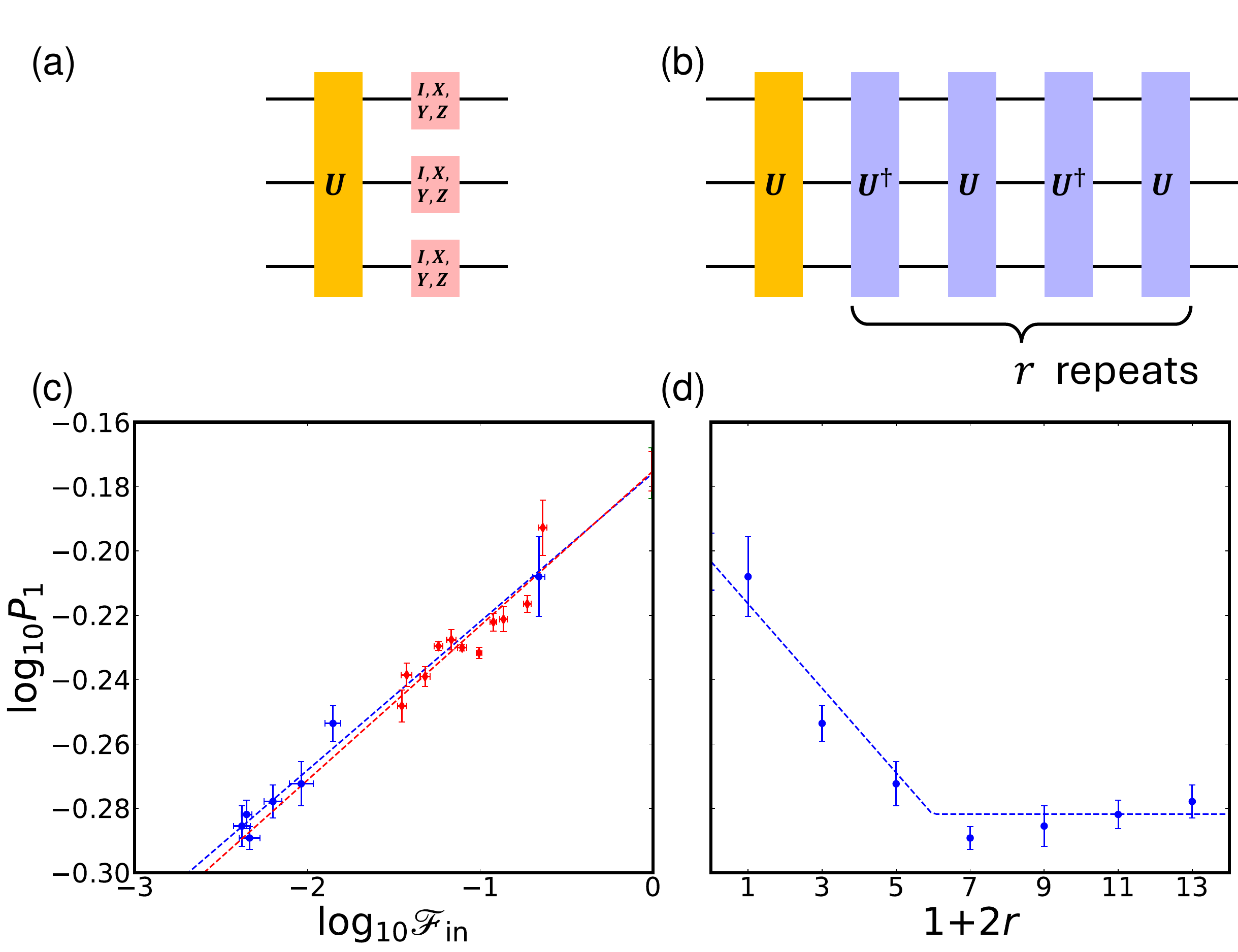}
\caption{\label{fig:EM} { Error mitigation ansatz.} (a)\&(b): Two different methods for boosting errors in the circuit: (a) adding single-qubit Pauli errors; (b) replacing $U$ with $U\left(U^\dagger U\right)^r$. (c) Error mitigation according to input fidelity. Linear regression suggests $P_1\approx67\pm 1\%$ in the limit of a perfect input state. The red and blue dots correspond to Monte Carlo sampling of single-qubit Pauli errors (averaged over $10^3$ noise realizations) and repeating $U^\dagger U$, respectively. (d) Error mitigation according to error rate in the second approach. Linear regression using data prior to saturation suggests $P_1\approx63\pm 1\%$ in the limit of a perfect input state. }
\end{center}
\end{figure}

To further enhance the performance of QCNN, we use error mitigation techniques to extrapolate to the noiseless limit. In particular, there is error from the state preparation circuit for $|\mathcal{S}_1\rangle$, due to the imperfect two-qubit gates that implement non-local swaps. Our error mitigation technique uses two different methods to boost error, which yields a relation between error rate and the overall performance that can be used for extrapolation. The first method involves randomly adding single-qubit Pauli gates to multi-qubit gates, shown in Fig.~\ref{fig:EM}(a).
The effect of such single-qubit errors can be simulated by Monte Carlo sampling.
The second method uses the transformation $U\rightarrow U\left(U^\dagger U\right)^r$, as shown in Fig. \ref{fig:EM}(b). In the noiseless situation, this is equivalent to applying a single unitary and hence has no effect on the result. On a noisy device, however, appending additional layers of $U^\dagger U$ increases the net error accumulated during the circuit evolution, and the performance becomes worse as the number of appended layers increase. Moreover, it can be directly implemented on the quantum device.

In Fig.~\ref{fig:EM}(c) and (d), we show results obtained using the two error mitigation methods discussed above. In Fig.~\ref{fig:EM}(c), we plot the success rate $P_1$ against the fidelity of the input state $\mathcal{F}_{\rm in}$, for different error rates via either adjusting the rate of single-qubit errors, or adjusting the number of appended layers $r$ in $(U^\dagger U)^r$. The method for estimating $\mathcal{F}_{\rm in}$ is detailed in Appendix~\ref{App:Exp}.
Regression analysis in the log-log scale suggests that the success rate of the QCNN, in the limit of a perfect input state $|\mathcal{S}_1\rangle$, reaches 67\%. In Fig.~\ref{fig:EM}(d), we instead plot the success rate $P_1$ against the number of appended layers $r$ in the second approach. By repeating the gates using $U\rightarrow U\left(U^\dagger U\right)^r$, the error is assumed to be $1+2r$ times that of $U$. The success probability decreases with increasing $r$ and eventually saturates due to the finite size of the Hilbert space. Regression analysis before saturation suggests that the success rate of the QCNN, in the absence of input errors, reaches 63\%.

Although experimental errors weaken the performance of QCNN, the classification signal is not completely drowned out by the noise.  Our experimental results demonstrate that QCNN can still achieve a good success rate in the presence of noise, which we recover by using error mitigation. In this work we mitigate the state preparation noise, while the QCNN circuit error mitigation will be deferred to future works dedicated to full experimentation.

\subsection{Generalizations to other models}
\begin{figure}[tb]
\begin{center}
\includegraphics[clip = true, width =\columnwidth]{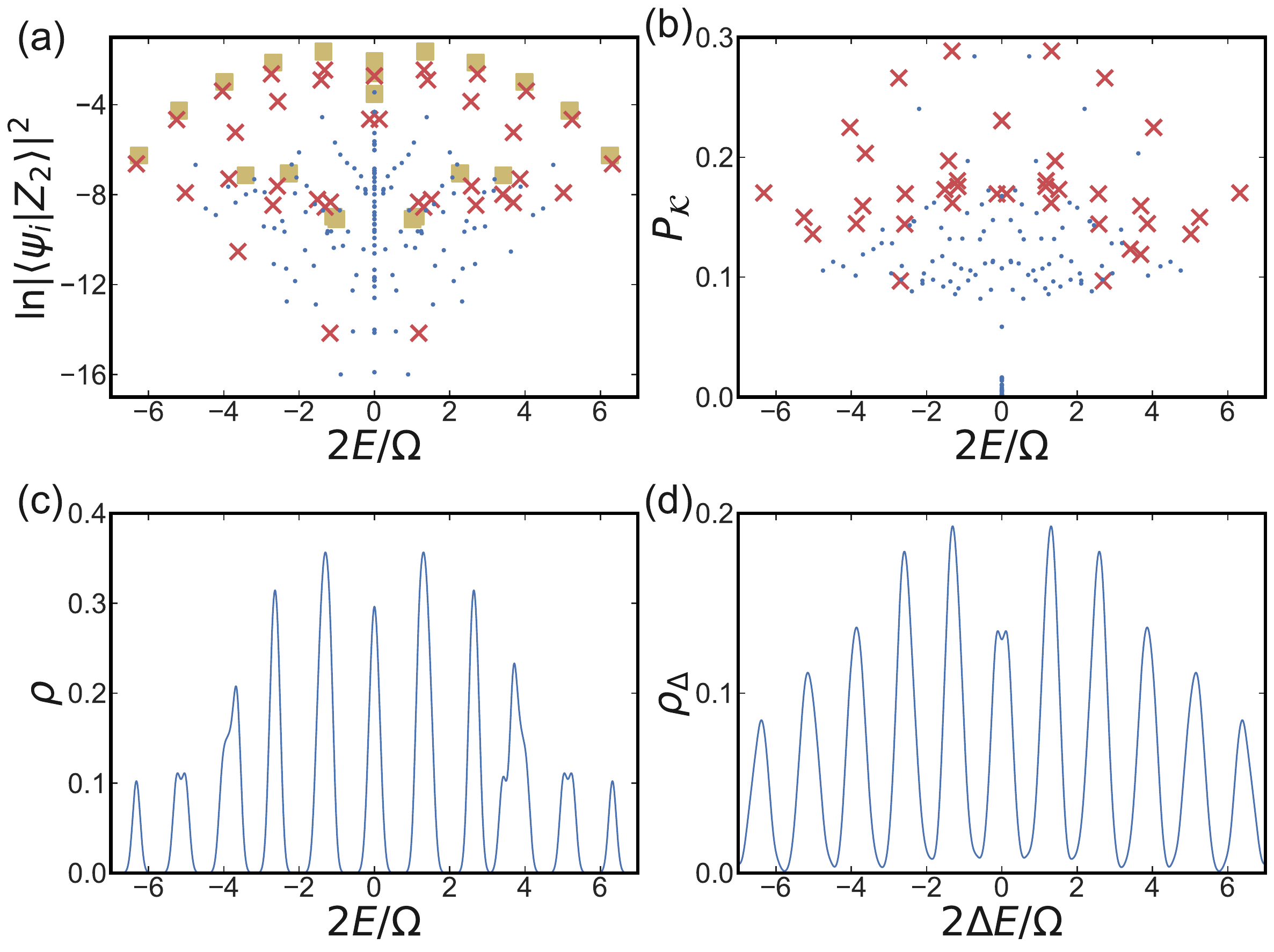}
\caption{\label{fig:PXP} { Additional marked states in the PXP model.} (a) Overlap between the eigenstates and the $|Z_2\rangle$ state, with red crosses marking the states identified by the QCNN. The yellow squares are the eigenstates of quasimodes in the symmetric subspace $\mathcal{K}$.(b) Weights of each eigenstates within the symmetric subspace $\mathcal{K}$. (c) Density of QCNN-identified states based in (a). (d) Density of energy differences between the QCNN-identified states in (a). In both (c) and (d), only states with non-zero overlap with the $|Z_{2}\rangle$ state are counted, with smoothing applied using Gaussian broadening of each points set to $0.1\Omega/2$.}
\end{center}
\end{figure}
In this section, we extend our QCNN-based approach to two additional models that host QMBSs. We begin with the PXP model, utilizing training data that include both analytically solvable and numerically identified scar states. Specifically, we incorporate the four exact scar states reported in Ref. \cite{Lin2019} and consider the states with the largest overlap with the $|Z_2\rangle$ across various energy windows as scar states \cite{Turner2018}.
Notably, the QCNN identifies many additional states as potentially non-thermal states, as shown in Fig. \ref{fig:PXP}(a). Some of these states exhibit smaller overlaps with $|Z_2\rangle$, remaining hidden within the chaotic spectrum. Inspired by Ref. \cite{Turner2021}, a symmetric subspace $\mathcal{K}$ can be constructed, which exhibits regular motion. The quasimodes within this subspace can be viewed as approximations of certain eigenstates. The subspace $\mathcal{K}$ is spanned by the basis states
\begin{eqnarray}
    |n_1,n_2\rangle=\frac{1}{\sqrt{\mathcal{N}_{n_1,n_2}}}\sum_{x\in(n_1,n_2)} |x\rangle ~,
\end{eqnarray}
where $x$ represents the binary configuration of the spin chain, with the constraint that neighboring spins cannot both be in the "1" state. Here, $n_1$ and $n_2$ denote the total number of "1"s at odd and even positions, respectively, and $\mathcal{N}_{n_1,n_2}$ is the normalization factor. The Hamiltonian in this subspace is expressed as $\langle n_1, n_2 | H_{\rm xorX} | n_1', n_2' \rangle$. The eigenstates of quasimodes in this subspace are shown as yellow squares in Fig. \ref{fig:PXP}(a), and these quasimodes closely align with some of the marked states.
In Fig. \ref{fig:PXP}(b), we plot the probability distribution of eigenstates within the subspace $\mathcal{K}$. Notably, the marked states have significant components within this subspace, demonstrating that the QCNN effectively learns hidden properties of the quasimodes without prior knowledge. Additionally, the QCNN not only identifies the top band in Fig. \ref{fig:PXP}(a), which is near the quasimodes, but also marks a second, lower band. This band has attracted considerable interest, though it still lacks a theoretical explanation.

The dynamics of these states are governed by their energy spectrum. The marked states exhibit energies similar to those of the $Z_2$ tower states. These states form distinct energy towers, as depicted in Fig. \ref{fig:PXP}(c). Their energies are approximately equally spaced, as shown in Fig. \ref{fig:PXP}(d), indicating that their linear superposition can lead to stable oscillations.

The perturbations in the PXP model exhibit distinct behaviors. In cases where perturbations enhance quantum many-body scars~\cite{Choi2019, Khemani2019, Desaules2024}, we observe a reduction in the number of non-thermal states identified by the QCNN. Under a uniform magnetic field perturbation, the QCNN additionally marks the second-highest energy band, alongside the top band, which is generally recognized as hosting scar states. Results for various perturbations are presented in Appendix~\ref{App:Pert}.

\begin{figure}[tb]
\begin{center}
\includegraphics[clip = true, width =\columnwidth]{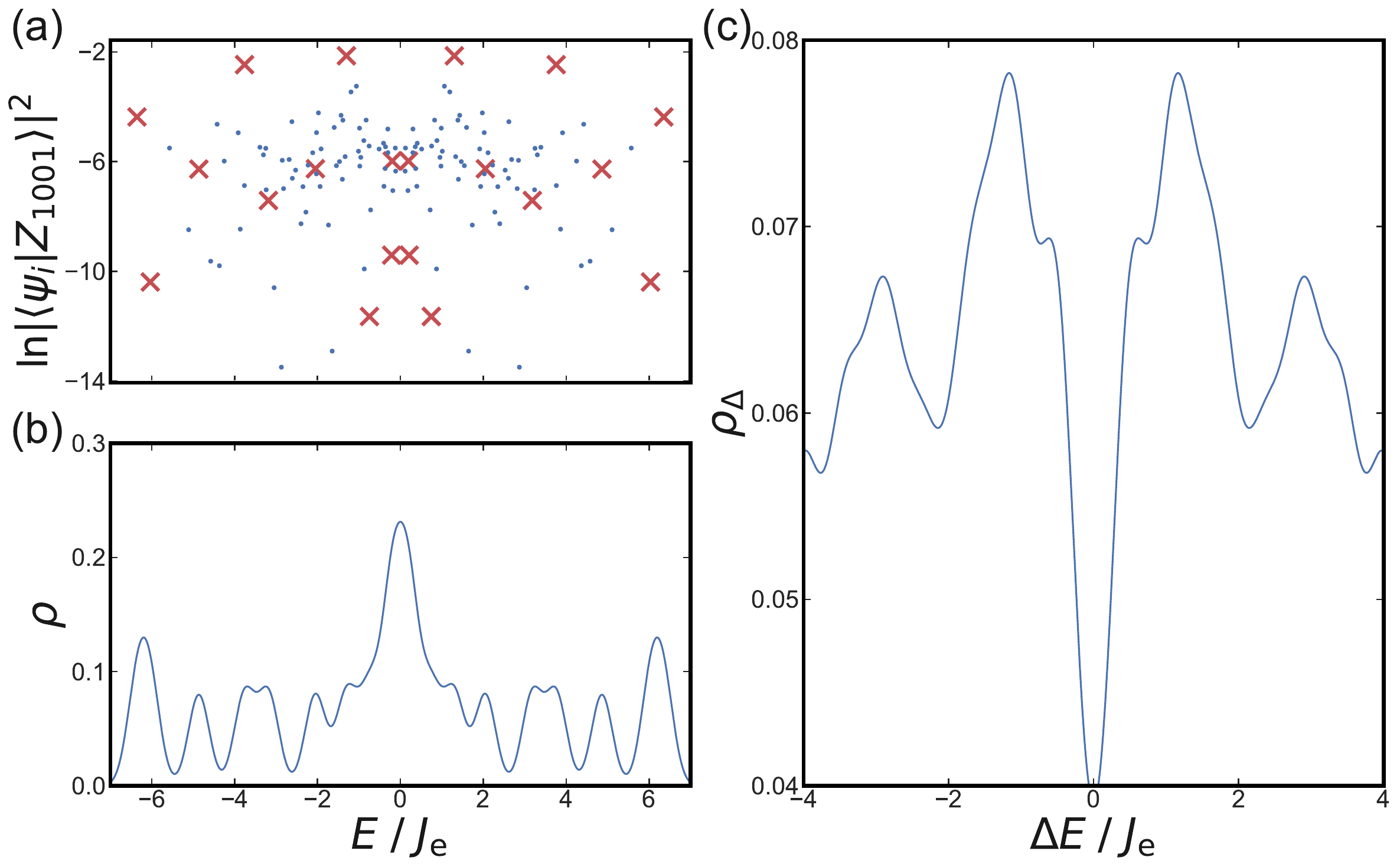}
\caption{\label{fig:Ising} { Additional marked states in the ar-coupling Ising SSH model.} (a) Overlap between the eigenstates and the $Z_{1001}$ state. The red crosses are states marked by QCNN. (b) Density of QCNN-identified states based in (a). (c) Density of energy differences between the QCNN-identified states in (a). In both (b) and (c), only states with non-zero overlap with the $|Z_{1001}\rangle$ state are counted, with smoothing applied using Gaussian broadening of each points set to $0.25J_{\rm e}$.}
\end{center}
\end{figure}

We also train the QCNN using the far-coupling Ising SSH model. The training data includes numerically solved scar states that exhibit significant overlaps with the $Z_{1001}$ state~\cite{ZhangPengfei2023}. Due to the smaller number of scar states in this model compared to others with the same number of qubits, the accuracy of the QCNN is reduced. Nonetheless, the QCNN identifies several additional states, as shown in Fig. \ref{fig:Ising}(a). Some of these states have smaller overlaps with the $Z_{1001}$ state. Although the peaks in the energy spectrum shown in Fig. \ref{fig:Ising}(b) appear mixed and unclear, we can clarify their behavior by examining the energy spectrum in Fig.~\ref{fig:Ising}(c).
In Fig.~\ref{fig:Ising}(c), the towers near $\Delta E=\pm 2.9J_{\rm e}$ are primarily contributed by the scar states, whereas the towers near $\Delta E=\pm 1.2J_{\rm e}$ are contributed by the additional states identified by QCNN. This suggests that the newly found non-thermal states exhibit a different oscillation frequency compared to the scar states.

We extend the QCNN approach to other models and discover additional states. The energy difference spectrum reveals that these identified states are non-thermal with relatively small dispersion. This demonstrates the QCNN's capability to identify non-thermal states across various models. Analytical understanding of these non-thermal states, similar to that presented for the xorX model in this work, is an interesting open problem.

\section{DISCUSSIONS}
Recent advances in the use of QML to investigate scar states have garnered considerable interest, resulting in a variety of related research. Here, we discuss differences between their studies and our approach. For instance, Ref.~\cite{Tomasz2022} introduces a general classifier using quantum variational autoencoders, where each eigenstate is assigned its own learning circuit. This approach can be computationally expensive, with costs scaling exponentially as Hilbert spaces grow in size. In contrast, our method employs a single circuit to learn the common properties of known scar states, enabling the discovery of other similar states. This makes our scheme more computationally feasible.
Meanwhile, Ref.~\cite{han2023meta} employs classical machine learning to detect single-body scar in billiard systems, given the classical description of eigenstates, but this methodology does not extend to many-body scars. Ref.~\cite{cao2024} uses classical machine learning and dimensionality reduction techniques to minimize the distance among scar states, but this can lead to loss of quantum information.
Although recent work shows that certain QCNNs can be simulated efficiently on classical systems~\cite{bermejo2024quantum}, our QCNN differs in two key ways: (1) Our QCNN is enhanced with additional universal layers in front of the original QCNN. We anticipate that using a small number of these layers (scaling logarithmically or linearly with system size) will make classical simulation hard. (2)  Preparing eigenstates of many-body systems remains a computationally challenging task. Scaling up system sizes will still rely on quantum hardware, complemented by techniques such as the variational eigensolver~\cite{tilly2022variational} for efficient eigenstate preparation.

\QZ{The convolutional layers in QCNNs are designed to detect local patterns and correlations in quantum states, making them particularly effective for identifying scar states, which exhibit sub-volume law entanglement entropy. Scar states are well-represented by matrix product states (MPS) \cite{Ho2019,Lin2019,Zhang2023}, which can be efficiently generated by tree-structured quantum circuits \cite{Malz2024}. Given that QCNNs have an inverse tree structure, they are inherently well-suited for capturing the characteristics of scar states.
Furthermore, the pooling operation in QCNNs parallels the coarse-graining step in the renormalization group (RG), simplifying the system while retaining essential information. Both QCNNs and RG employ multi-scale analysis to extract key features, albeit in different contexts \cite{liu2023model}. A deeper exploration of this correspondence and its underlying mechanisms remains an open avenue for further investigation.}

In conclusion, the QCNN trained on scar states effectively identifies additional non-thermal states. Some of them primarily occupy a small fraction of the Hilbert space and can be approximately described as spin-wave modes of various quasiparticles. We construct effective Hamiltonians based on this framework, capturing key features of the exact many-body eigenstates. Additionally, we validate our approach on a quantum device, achieving a notable success rate with the use of error mitigation techniques. This study highlights the potential of QCNNs in uncovering hidden non-thermal states within the many-body spectrum, paving the way for future research into more complex quantum systems and their non-thermal behavior.

\section{METHODS}
\subsection{Enhancement of Quantum Convolutional Neural Network}
Our approach utilizes QML by feeding known scar states as the training data set. However, scars are rare, as their number grows linearly with the number of qubits, whereas  the Hilbert space expands exponentially. To address this limitation, we include superpositions of QMBS states in our training dataset, enhancing the model's ability to learn from a small set of scar states.

For efficient training, we adopt QCNN as the quantum circuit ansatz. Its translational invariant gate architecture is well-suited for handling translational invariant Hamiltonians. Additionally, scar states can often be expressed as matrix-product states~\cite{Lin2019,Zhang2023}, which QCNN can learn efficiently.

As illustrated in Fig.~\ref{fig:setup}, a QCNN is composed of convolution layers, pooling layers and a fully connected layer~\cite{Lukin2019,Caro2022}. Each convolutional layer consists of brick-wall two-qubit gates with identical parameters, reducing the number of parameters and facilitating more efficient classical optimization \cite{Cerezo2021}.
These gates act as a quantum kernel, convolving the wave function through each layer. Furthermore, the locality alleviates the barren plateau problem \cite{Cerezo2021}. The pooling layers measure a subset of qubits after the quantum gates, reducing the number of operations needed in subsequent steps. This not only makes the quantum circuit shallower but also reduces noise. The fully connected layer then consolidates the information from the remaining qubits at the end of the process.

\begin{figure}[tb]
\begin{center}
\includegraphics[clip = true, width =\columnwidth]{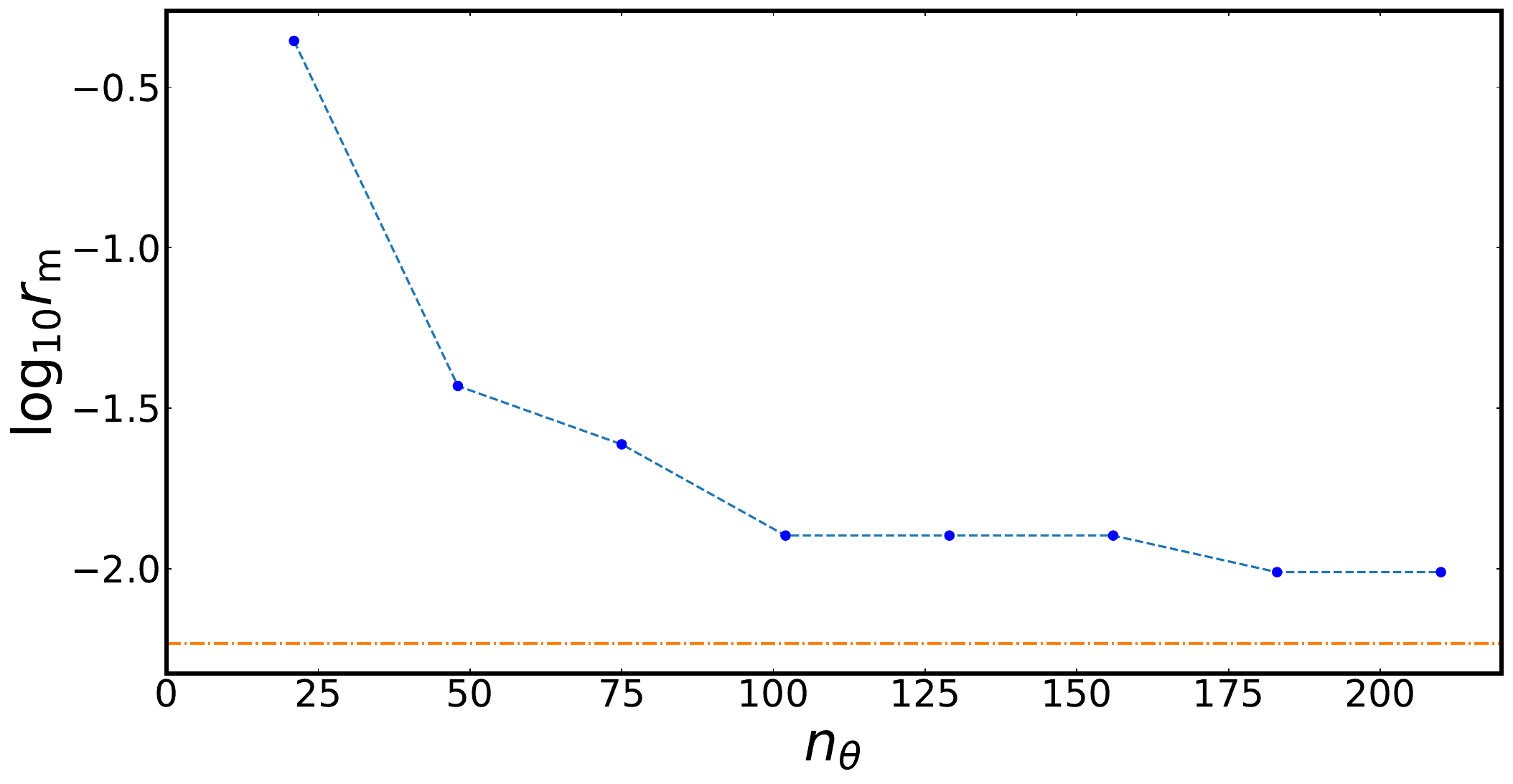}
\caption{\label{fig:LossPara}
Training the xorX model with varying numbers of trainable gate parameters, denoted as $n_\theta$. The rate of marked states $r_m$ is the ratio of marked states to the total dimension of the Hilbert space. Each point represents a distinct set of circuit parameters and a corresponding training iteration. The orange dashed line indicates the rate of marked (scar) states in the input data. The system consists of 12 spins. The number of trainable parameters is given by $n_\theta=12+3(9n_l+3)$, where $n_l$ is the number of convolutional layers preceding the pooling layers.}
\end{center}
\end{figure}

We further enhance the QCNN to improve its ability to classify quantum states with greater precision. Specifically, we introduce additional convolution layers after each existing convolution layer to enable the network to capture correlations over longer distances, analogous to the effect of using larger kernels in classical convolutional neural networks. The performance with different numbers of convolutional layers is shown in Fig. \ref{fig:LossPara}. To tackle more complex classification tasks, we introduce a few general layers of universal quantum gates to preprocess the input data. This adjustment helps handle finite-size effect and outliers which breaks translational-symmetry, such as disorders or boundaries.
Furthermore, we incorporate an ancillary qubit initialized in the $|0\rangle$ state, analogous to the zero padding at the boundary pixels in classical convolutional neural networks. This allows us to perform classification operations without altering the size of the Hilbert space (see Appendix~\ref{App:QCNN}).

When the circuit has too few layers, its limited discriminative power causes various states to become indistinguishable. As the number of parameterized gates increases, the classification error gradually decreases. This improvement may lead to overfitting in an over-parameterized circuit. The rate of marked states (i.e., those identified as scar states by the QCNN) tends to align with the rate in the input data, as shown in Fig. \ref{fig:LossPara}. Classification becomes saturated when the number of parameters approaches the size of the Hilbert space. The states identified by the circuit will transition between exact scar and chaotic states, with this transition expected to be continuous, akin to the classical case. In other words, the number of parameters controls the proximity of the regime to the exact scar states that are marked \cite{cao2024, Eswarathasan2017}. Our study focuses on this transition to identify previously unknown non-thermal states with scar-like characteristics.

\subsection{Training with Scar States}
To train the enhanced QCNN, we use training data generated from known QMBS states. Specifically, for a system with known scar states, we label these known scar states and their superpositions as $y_i=1$, while labeling other eigenstates and their superpositions as $y_i=0$. We then randomly select an equal number of states from each label category and train the enhanced QCNN using gradient descent to minimize the loss function \cite{Caro2022}
\begin{eqnarray}
  \mathcal{L} &=&\frac{1}{d} \sum_{i=0}^{d-1}\left|y_i-q_i \right| ~,
\end{eqnarray}
where $d$ is the total number of input states, and $q_i$ denotes the probability of the output qubit being in $\ket{1}$ state. This $q_i$ also corresponds to the single-shot measurement accuracy for identifying exact scar states. The enhanced QCNN is expected to distinguish scar states from thermal states after sufficient training with enough number of layers. After training, we evaluate the QCNN's performance by testing it on the eigenstates of the Hamiltonian.
For convenience, we consider an eigenstate to be ``marked" if the output
$q_i$ exceeds 50\%, matching the rate of scar states in the training set. Experimentally, this classification is achieved through a majority vote across multiple trials and measurements \cite{Harry2022}. In this context, achieving zero loss indicates perfect classification, although zero classification error does not necessarily imply the loss function fully decays to zero. The loss function also reflects the model's robustness to noise.

\begin{acknowledgements}

The authors acknowledge discussions with Zlatko Papic on quasi-modes in the PXP model.
J.-J.F., B.Z. and Q.Z. acknowledge support from University of Southern California. Z.-C. Y. acknowledges support from Peking University. This research was conducted using IBM Quantum Systems provided through USC’s IBM Quantum Innovation Center.

J.-J.F. and Q.Z. proposed the study of scar states using quantum machine learning. Q.Z. designed the framework of quantum machine learning. J.-J.F. performed numerical analyses and experiments, with support from B.Z., under the supervision of Q.Z. J.-J.F. derived the analytical theory. Z.-C.Y. provided insights on quantum many body scars. J.-J.F. and Q.Z. wrote the manuscript, with inputs and contributions from all authors.

\end{acknowledgements}


\begin{thebibliography}{58}%
\makeatletter
\providecommand \@ifxundefined [1]{%
 \@ifx{#1\undefined}
}%
\providecommand \@ifnum [1]{%
 \ifnum #1\expandafter \@firstoftwo
 \else \expandafter \@secondoftwo
 \fi
}%
\providecommand \@ifx [1]{%
 \ifx #1\expandafter \@firstoftwo
 \else \expandafter \@secondoftwo
 \fi
}%
\providecommand \natexlab [1]{#1}%
\providecommand \enquote  [1]{``#1''}%
\providecommand \bibnamefont  [1]{#1}%
\providecommand \bibfnamefont [1]{#1}%
\providecommand \citenamefont [1]{#1}%
\providecommand \href@noop [0]{\@secondoftwo}%
\providecommand \href [0]{\begingroup \@sanitize@url \@href}%
\providecommand \@href[1]{\@@startlink{#1}\@@href}%
\providecommand \@@href[1]{\endgroup#1\@@endlink}%
\providecommand \@sanitize@url [0]{\catcode `\\12\catcode `\$12\catcode
  `\&12\catcode `\#12\catcode `\^12\catcode `\_12\catcode `\%12\relax}%
\providecommand \@@startlink[1]{}%
\providecommand \@@endlink[0]{}%
\providecommand \url  [0]{\begingroup\@sanitize@url \@url }%
\providecommand \@url [1]{\endgroup\@href {#1}{\urlprefix }}%
\providecommand \urlprefix  [0]{URL }%
\providecommand \Eprint [0]{\href }%
\providecommand \doibase [0]{https://doi.org/}%
\providecommand \selectlanguage [0]{\@gobble}%
\providecommand \bibinfo  [0]{\@secondoftwo}%
\providecommand \bibfield  [0]{\@secondoftwo}%
\providecommand \translation [1]{[#1]}%
\providecommand \BibitemOpen [0]{}%
\providecommand \bibitemStop [0]{}%
\providecommand \bibitemNoStop [0]{.\EOS\space}%
\providecommand \EOS [0]{\spacefactor3000\relax}%
\providecommand \BibitemShut  [1]{\csname bibitem#1\endcsname}%
\let\auto@bib@innerbib\@empty
\bibitem [{\citenamefont {von Neumann}(2010)}]{von2010proof}%
  \BibitemOpen
  \bibfield  {author} {\bibinfo {author} {\bibfnamefont {J.}~\bibnamefont {von
  Neumann}},\ }\bibfield  {title} {\bibinfo {title} {Proof of the ergodic
  theorem and the h-theorem in quantum mechanics: Translation of: Beweis des
  ergodensatzes und des h-theorems in der neuen mechanik},\ }\href@noop {}
  {\bibfield  {journal} {\bibinfo  {journal} {Eur. Phys. J. H}\ }\textbf
  {\bibinfo {volume} {35}},\ \bibinfo {pages} {201} (\bibinfo {year}
  {2010})}\BibitemShut {NoStop}%
\bibitem [{\citenamefont {Reimann}(2008)}]{reimann2008foundation}%
  \BibitemOpen
  \bibfield  {author} {\bibinfo {author} {\bibfnamefont {P.}~\bibnamefont
  {Reimann}},\ }\bibfield  {title} {\bibinfo {title} {Foundation of statistical
  mechanics under experimentally realistic conditions},\ }\href@noop {}
  {\bibfield  {journal} {\bibinfo  {journal} {Phys. Rev. Lett.}\ }\textbf
  {\bibinfo {volume} {101}},\ \bibinfo {pages} {190403} (\bibinfo {year}
  {2008})}\BibitemShut {NoStop}%
\bibitem [{\citenamefont {Zhang}\ \emph {et~al.}(2016)\citenamefont {Zhang},
  \citenamefont {Quan},\ and\ \citenamefont {Wu}}]{Zhang2016}%
  \BibitemOpen
  \bibfield  {author} {\bibinfo {author} {\bibfnamefont {D.}~\bibnamefont
  {Zhang}}, \bibinfo {author} {\bibfnamefont {H.~T.}\ \bibnamefont {Quan}},\
  and\ \bibinfo {author} {\bibfnamefont {B.}~\bibnamefont {Wu}},\ }\bibfield
  {title} {\bibinfo {title} {Ergodicity and mixing in quantum dynamics},\
  }\href {https://doi.org/10.1103/PhysRevE.94.022150} {\bibfield  {journal}
  {\bibinfo  {journal} {Phys. Rev. E}\ }\textbf {\bibinfo {volume} {94}},\
  \bibinfo {pages} {022150} (\bibinfo {year} {2016})}\BibitemShut {NoStop}%
\bibitem [{\citenamefont {Richens}\ and\ \citenamefont
  {Berry}(1981)}]{richens1981pseudointegrable}%
  \BibitemOpen
  \bibfield  {author} {\bibinfo {author} {\bibfnamefont {P.}~\bibnamefont
  {Richens}}\ and\ \bibinfo {author} {\bibfnamefont {M.}~\bibnamefont
  {Berry}},\ }\bibfield  {title} {\bibinfo {title} {Pseudointegrable systems in
  classical and quantum mechanics},\ }\href
  {https://doi.org/https://doi.org/10.1016/0167-2789(81)90024-5} {\bibfield
  {journal} {\bibinfo  {journal} {Physica D: Nonlinear Phenomena}\ }\textbf
  {\bibinfo {volume} {2}},\ \bibinfo {pages} {495} (\bibinfo {year}
  {1981})}\BibitemShut {NoStop}%
\bibitem [{\citenamefont {Dragović}\ and\ \citenamefont
  {Radnović}(2015)}]{dragovic2015periods}%
  \BibitemOpen
  \bibfield  {author} {\bibinfo {author} {\bibfnamefont {V.}~\bibnamefont
  {Dragović}}\ and\ \bibinfo {author} {\bibfnamefont {M.}~\bibnamefont
  {Radnović}},\ }\bibfield  {title} {\bibinfo {title} {Periods of
  pseudo-integrable billiards},\ }\href
  {https://doi.org/10.1007/s40598-014-0004-0} {\bibfield  {journal} {\bibinfo
  {journal} {Arnold Mathematical Journal}\ }\textbf {\bibinfo {volume} {1}},\
  \bibinfo {pages} {69} (\bibinfo {year} {2015})}\BibitemShut {NoStop}%
\bibitem [{\citenamefont {Srednicki}(1994)}]{srednicki1994chaos}%
  \BibitemOpen
  \bibfield  {author} {\bibinfo {author} {\bibfnamefont {M.}~\bibnamefont
  {Srednicki}},\ }\bibfield  {title} {\bibinfo {title} {Chaos and quantum
  thermalization},\ }\href {https://doi.org/10.1103/PhysRevE.50.888} {\bibfield
   {journal} {\bibinfo  {journal} {Phys. Rev. E}\ }\textbf {\bibinfo {volume}
  {50}},\ \bibinfo {pages} {888} (\bibinfo {year} {1994})}\BibitemShut
  {NoStop}%
\bibitem [{\citenamefont {Deutsch}(1991)}]{deutsch1991}%
  \BibitemOpen
  \bibfield  {author} {\bibinfo {author} {\bibfnamefont {J.~M.}\ \bibnamefont
  {Deutsch}},\ }\bibfield  {title} {\bibinfo {title} {Quantum statistical
  mechanics in a closed system},\ }\href
  {https://doi.org/10.1103/PhysRevA.43.2046} {\bibfield  {journal} {\bibinfo
  {journal} {Phys. Rev. A}\ }\textbf {\bibinfo {volume} {43}},\ \bibinfo
  {pages} {2046} (\bibinfo {year} {1991})}\BibitemShut {NoStop}%
\bibitem [{\citenamefont {Rigol}\ \emph {et~al.}(2008)\citenamefont {Rigol},
  \citenamefont {Dunjko},\ and\ \citenamefont
  {Olshanii}}]{rigol2008thermalization}%
  \BibitemOpen
  \bibfield  {author} {\bibinfo {author} {\bibfnamefont {M.}~\bibnamefont
  {Rigol}}, \bibinfo {author} {\bibfnamefont {V.}~\bibnamefont {Dunjko}},\ and\
  \bibinfo {author} {\bibfnamefont {M.}~\bibnamefont {Olshanii}},\ }\bibfield
  {title} {\bibinfo {title} {Thermalization and its mechanism for generic
  isolated quantum systems},\ }\href {https://doi.org/10.1038/nature06838}
  {\bibfield  {journal} {\bibinfo  {journal} {Nature}\ }\textbf {\bibinfo
  {volume} {452}},\ \bibinfo {pages} {854} (\bibinfo {year}
  {2008})}\BibitemShut {NoStop}%
\bibitem [{\citenamefont {Popescu}\ \emph {et~al.}(2006)\citenamefont
  {Popescu}, \citenamefont {Short},\ and\ \citenamefont
  {Winter}}]{popescu2006entanglement}%
  \BibitemOpen
  \bibfield  {author} {\bibinfo {author} {\bibfnamefont {S.}~\bibnamefont
  {Popescu}}, \bibinfo {author} {\bibfnamefont {A.~J.}\ \bibnamefont {Short}},\
  and\ \bibinfo {author} {\bibfnamefont {A.}~\bibnamefont {Winter}},\
  }\bibfield  {title} {\bibinfo {title} {Entanglement and the foundations of
  statistical mechanics},\ }\href {https://doi.org/10.1038/nphys444} {\bibfield
   {journal} {\bibinfo  {journal} {Nature Physics}\ }\textbf {\bibinfo {volume}
  {2}},\ \bibinfo {pages} {754} (\bibinfo {year} {2006})}\BibitemShut {NoStop}%
\bibitem [{\citenamefont {Kim}\ \emph {et~al.}(2014)\citenamefont {Kim},
  \citenamefont {Ikeda},\ and\ \citenamefont {Huse}}]{Kim2014}%
  \BibitemOpen
  \bibfield  {author} {\bibinfo {author} {\bibfnamefont {H.}~\bibnamefont
  {Kim}}, \bibinfo {author} {\bibfnamefont {T.~N.}\ \bibnamefont {Ikeda}},\
  and\ \bibinfo {author} {\bibfnamefont {D.~A.}\ \bibnamefont {Huse}},\
  }\bibfield  {title} {\bibinfo {title} {Testing whether all eigenstates obey
  the eigenstate thermalization hypothesis},\ }\href
  {https://doi.org/10.1103/PhysRevE.90.052105} {\bibfield  {journal} {\bibinfo
  {journal} {Phys. Rev. E}\ }\textbf {\bibinfo {volume} {90}},\ \bibinfo
  {pages} {052105} (\bibinfo {year} {2014})}\BibitemShut {NoStop}%
\bibitem [{\citenamefont {Short}\ and\ \citenamefont
  {Farrelly}(2012)}]{short2012quantum}%
  \BibitemOpen
  \bibfield  {author} {\bibinfo {author} {\bibfnamefont {A.~J.}\ \bibnamefont
  {Short}}\ and\ \bibinfo {author} {\bibfnamefont {T.~C.}\ \bibnamefont
  {Farrelly}},\ }\bibfield  {title} {\bibinfo {title} {Quantum equilibration in
  finite time},\ }\href@noop {} {\bibfield  {journal} {\bibinfo  {journal} {New
  Journal of Physics}\ }\textbf {\bibinfo {volume} {14}},\ \bibinfo {pages}
  {013063} (\bibinfo {year} {2012})}\BibitemShut {NoStop}%
\bibitem [{\citenamefont {Serbyn}\ \emph {et~al.}(2021)\citenamefont {Serbyn},
  \citenamefont {Abanin},\ and\ \citenamefont {Papić}}]{Serbyn2021}%
  \BibitemOpen
  \bibfield  {author} {\bibinfo {author} {\bibfnamefont {M.}~\bibnamefont
  {Serbyn}}, \bibinfo {author} {\bibfnamefont {D.~A.}\ \bibnamefont {Abanin}},\
  and\ \bibinfo {author} {\bibfnamefont {Z.}~\bibnamefont {Papić}},\
  }\bibfield  {title} {\bibinfo {title} {Quantum many-body scars and weak
  breaking of ergodicity},\ }\href {https://doi.org/10.1038/s41567-021-01230-2}
  {\bibfield  {journal} {\bibinfo  {journal} {Nature Physics}\ }\textbf
  {\bibinfo {volume} {17}},\ \bibinfo {pages} {675} (\bibinfo {year}
  {2021})}\BibitemShut {NoStop}%
\bibitem [{\citenamefont {Moudgalya}\ \emph {et~al.}(2022)\citenamefont
  {Moudgalya}, \citenamefont {Bernevig},\ and\ \citenamefont
  {Regnault}}]{moudgalya2022quantum}%
  \BibitemOpen
  \bibfield  {author} {\bibinfo {author} {\bibfnamefont {S.}~\bibnamefont
  {Moudgalya}}, \bibinfo {author} {\bibfnamefont {B.~A.}\ \bibnamefont
  {Bernevig}},\ and\ \bibinfo {author} {\bibfnamefont {N.}~\bibnamefont
  {Regnault}},\ }\bibfield  {title} {\bibinfo {title} {Quantum many-body scars
  and hilbert space fragmentation: a review of exact results},\ }\href
  {https://doi.org/10.1088/1361-6633/ac73a0} {\bibfield  {journal} {\bibinfo
  {journal} {Reports on Progress in Physics}\ }\textbf {\bibinfo {volume}
  {85}},\ \bibinfo {pages} {086501} (\bibinfo {year} {2022})}\BibitemShut
  {NoStop}%
\bibitem [{\citenamefont {Dong}\ \emph {et~al.}(2023)\citenamefont {Dong},
  \citenamefont {Desaules}, \citenamefont {Gao}, \citenamefont {Wang},
  \citenamefont {Guo}, \citenamefont {Chen}, \citenamefont {Zou}, \citenamefont
  {Jin}, \citenamefont {Zhu}, \citenamefont {Zhang}, \citenamefont {Li},
  \citenamefont {Wang}, \citenamefont {Guo}, \citenamefont {Zhang},
  \citenamefont {Ying},\ and\ \citenamefont {Papić}}]{Hang2023}%
  \BibitemOpen
  \bibfield  {author} {\bibinfo {author} {\bibfnamefont {H.}~\bibnamefont
  {Dong}}, \bibinfo {author} {\bibfnamefont {J.-Y.}\ \bibnamefont {Desaules}},
  \bibinfo {author} {\bibfnamefont {Y.}~\bibnamefont {Gao}}, \bibinfo {author}
  {\bibfnamefont {N.}~\bibnamefont {Wang}}, \bibinfo {author} {\bibfnamefont
  {Z.}~\bibnamefont {Guo}}, \bibinfo {author} {\bibfnamefont {J.}~\bibnamefont
  {Chen}}, \bibinfo {author} {\bibfnamefont {Y.}~\bibnamefont {Zou}}, \bibinfo
  {author} {\bibfnamefont {F.}~\bibnamefont {Jin}}, \bibinfo {author}
  {\bibfnamefont {X.}~\bibnamefont {Zhu}}, \bibinfo {author} {\bibfnamefont
  {P.}~\bibnamefont {Zhang}}, \bibinfo {author} {\bibfnamefont
  {H.}~\bibnamefont {Li}}, \bibinfo {author} {\bibfnamefont {Z.}~\bibnamefont
  {Wang}}, \bibinfo {author} {\bibfnamefont {Q.}~\bibnamefont {Guo}}, \bibinfo
  {author} {\bibfnamefont {J.}~\bibnamefont {Zhang}}, \bibinfo {author}
  {\bibfnamefont {L.}~\bibnamefont {Ying}},\ and\ \bibinfo {author}
  {\bibfnamefont {Z.}~\bibnamefont {Papić}},\ }\bibfield  {title} {\bibinfo
  {title} {Disorder-tunable entanglement at infinite temperature},\ }\href
  {https://doi.org/10.1126/sciadv.adj3822} {\bibfield  {journal} {\bibinfo
  {journal} {Science Advances}\ }\textbf {\bibinfo {volume} {9}},\ \bibinfo
  {pages} {eadj3822} (\bibinfo {year} {2023})},\ \Eprint
  {https://arxiv.org/abs/https://www.science.org/doi/pdf/10.1126/sciadv.adj3822}
  {https://www.science.org/doi/pdf/10.1126/sciadv.adj3822} \BibitemShut
  {NoStop}%
\bibitem [{\citenamefont {Evrard}\ \emph {et~al.}(2024)\citenamefont {Evrard},
  \citenamefont {Pizzi}, \citenamefont {Mistakidis},\ and\ \citenamefont
  {Dag}}]{Evrard2024}%
  \BibitemOpen
  \bibfield  {author} {\bibinfo {author} {\bibfnamefont {B.}~\bibnamefont
  {Evrard}}, \bibinfo {author} {\bibfnamefont {A.}~\bibnamefont {Pizzi}},
  \bibinfo {author} {\bibfnamefont {S.~I.}\ \bibnamefont {Mistakidis}},\ and\
  \bibinfo {author} {\bibfnamefont {C.~B.}\ \bibnamefont {Dag}},\ }\bibfield
  {title} {\bibinfo {title} {Quantum scars and regular eigenstates in a chaotic
  spinor condensate},\ }\href {https://doi.org/10.1103/PhysRevLett.132.020401}
  {\bibfield  {journal} {\bibinfo  {journal} {Phys. Rev. Lett.}\ }\textbf
  {\bibinfo {volume} {132}},\ \bibinfo {pages} {020401} (\bibinfo {year}
  {2024})}\BibitemShut {NoStop}%
\bibitem [{\citenamefont {Bernien}\ \emph {et~al.}(2017)\citenamefont
  {Bernien}, \citenamefont {Schwartz}, \citenamefont {Keesling}, \citenamefont
  {Levine}, \citenamefont {Omran}, \citenamefont {Pichler}, \citenamefont
  {Choi}, \citenamefont {Zibrov}, \citenamefont {Endres}, \citenamefont
  {Greiner}, \citenamefont {Vuletić},\ and\ \citenamefont
  {Lukin}}]{bernien2017probing}%
  \BibitemOpen
  \bibfield  {author} {\bibinfo {author} {\bibfnamefont {H.}~\bibnamefont
  {Bernien}}, \bibinfo {author} {\bibfnamefont {S.}~\bibnamefont {Schwartz}},
  \bibinfo {author} {\bibfnamefont {A.}~\bibnamefont {Keesling}}, \bibinfo
  {author} {\bibfnamefont {H.}~\bibnamefont {Levine}}, \bibinfo {author}
  {\bibfnamefont {A.}~\bibnamefont {Omran}}, \bibinfo {author} {\bibfnamefont
  {H.}~\bibnamefont {Pichler}}, \bibinfo {author} {\bibfnamefont
  {S.}~\bibnamefont {Choi}}, \bibinfo {author} {\bibfnamefont {A.~S.}\
  \bibnamefont {Zibrov}}, \bibinfo {author} {\bibfnamefont {M.}~\bibnamefont
  {Endres}}, \bibinfo {author} {\bibfnamefont {M.}~\bibnamefont {Greiner}},
  \bibinfo {author} {\bibfnamefont {V.}~\bibnamefont {Vuletić}},\ and\
  \bibinfo {author} {\bibfnamefont {M.~D.}\ \bibnamefont {Lukin}},\ }\bibfield
  {title} {\bibinfo {title} {Probing many-body dynamics on a 51-atom quantum
  simulator},\ }\href {https://doi.org/10.1038/nature24622} {\bibfield
  {journal} {\bibinfo  {journal} {Nature}\ }\textbf {\bibinfo {volume} {551}},\
  \bibinfo {pages} {579} (\bibinfo {year} {2017})}\BibitemShut {NoStop}%
\bibitem [{\citenamefont {Iadecola}\ and\ \citenamefont
  {Schecter}(2020)}]{Iadecola2020}%
  \BibitemOpen
  \bibfield  {author} {\bibinfo {author} {\bibfnamefont {T.}~\bibnamefont
  {Iadecola}}\ and\ \bibinfo {author} {\bibfnamefont {M.}~\bibnamefont
  {Schecter}},\ }\bibfield  {title} {\bibinfo {title} {Quantum many-body scar
  states with emergent kinetic constraints and finite-entanglement revivals},\
  }\href {https://doi.org/10.1103/PhysRevB.101.024306} {\bibfield  {journal}
  {\bibinfo  {journal} {Phys. Rev. B}\ }\textbf {\bibinfo {volume} {101}},\
  \bibinfo {pages} {024306} (\bibinfo {year} {2020})}\BibitemShut {NoStop}%
\bibitem [{\citenamefont {Wang}\ \emph
  {et~al.}(2021{\natexlab{a}})\citenamefont {Wang}, \citenamefont {Feng},\ and\
  \citenamefont {Wu}}]{Wang2021}%
  \BibitemOpen
  \bibfield  {author} {\bibinfo {author} {\bibfnamefont {Z.}~\bibnamefont
  {Wang}}, \bibinfo {author} {\bibfnamefont {J.}~\bibnamefont {Feng}},\ and\
  \bibinfo {author} {\bibfnamefont {B.}~\bibnamefont {Wu}},\ }\bibfield
  {title} {\bibinfo {title} {Microscope for quantum dynamics with planck cell
  resolution},\ }\href {https://doi.org/10.1103/PhysRevResearch.3.033239}
  {\bibfield  {journal} {\bibinfo  {journal} {Phys. Rev. Res.}\ }\textbf
  {\bibinfo {volume} {3}},\ \bibinfo {pages} {033239} (\bibinfo {year}
  {2021}{\natexlab{a}})}\BibitemShut {NoStop}%
\bibitem [{\citenamefont {Eswarathasan}\ and\ \citenamefont
  {Nonnenmacher}(2017)}]{Eswarathasan2017}%
  \BibitemOpen
  \bibfield  {author} {\bibinfo {author} {\bibfnamefont {S.}~\bibnamefont
  {Eswarathasan}}\ and\ \bibinfo {author} {\bibfnamefont {S.}~\bibnamefont
  {Nonnenmacher}},\ }\bibfield  {title} {\bibinfo {title} {Strong scarring of
  logarithmic quasimodes},\ }\href {https://doi.org/10.5802/aif.3137}
  {\bibfield  {journal} {\bibinfo  {journal} {Annales de l'Institut Fourier}\
  }\textbf {\bibinfo {volume} {67}},\ \bibinfo {pages} {2307} (\bibinfo {year}
  {2017})}\BibitemShut {NoStop}%
\bibitem [{\citenamefont {Kong}\ \emph {et~al.}(2024)\citenamefont {Kong},
  \citenamefont {Gong},\ and\ \citenamefont {Wu}}]{Kong2024}%
  \BibitemOpen
  \bibfield  {author} {\bibinfo {author} {\bibfnamefont {L.}~\bibnamefont
  {Kong}}, \bibinfo {author} {\bibfnamefont {Z.}~\bibnamefont {Gong}},\ and\
  \bibinfo {author} {\bibfnamefont {B.}~\bibnamefont {Wu}},\ }\bibfield
  {title} {\bibinfo {title} {Quantum dynamical tunneling breaks classical
  conserved quantities},\ }\href {https://doi.org/10.1103/PhysRevE.109.054113}
  {\bibfield  {journal} {\bibinfo  {journal} {Phys. Rev. E}\ }\textbf {\bibinfo
  {volume} {109}},\ \bibinfo {pages} {054113} (\bibinfo {year}
  {2024})}\BibitemShut {NoStop}%
\bibitem [{\citenamefont {Surace}\ and\ \citenamefont
  {Motrunich}(2023)}]{Surace2023}%
  \BibitemOpen
  \bibfield  {author} {\bibinfo {author} {\bibfnamefont {F.~M.}\ \bibnamefont
  {Surace}}\ and\ \bibinfo {author} {\bibfnamefont {O.}~\bibnamefont
  {Motrunich}},\ }\bibfield  {title} {\bibinfo {title} {Weak integrability
  breaking perturbations of integrable models},\ }\href
  {https://doi.org/10.1103/PhysRevResearch.5.043019} {\bibfield  {journal}
  {\bibinfo  {journal} {Phys. Rev. Res.}\ }\textbf {\bibinfo {volume} {5}},\
  \bibinfo {pages} {043019} (\bibinfo {year} {2023})}\BibitemShut {NoStop}%
\bibitem [{\citenamefont {Ivanov}\ and\ \citenamefont
  {Motrunich}(2025)}]{ivanov2024volumepxp}%
  \BibitemOpen
  \bibfield  {author} {\bibinfo {author} {\bibfnamefont {A.~N.}\ \bibnamefont
  {Ivanov}}\ and\ \bibinfo {author} {\bibfnamefont {O.~I.}\ \bibnamefont
  {Motrunich}},\ }\bibfield  {title} {\bibinfo {title} {Volume-entangled exact
  scar states in the pxp and related models in any dimension},\ }\href
  {https://doi.org/10.1103/PhysRevLett.134.050403} {\bibfield  {journal}
  {\bibinfo  {journal} {Phys. Rev. Lett.}\ }\textbf {\bibinfo {volume} {134}},\
  \bibinfo {pages} {050403} (\bibinfo {year} {2025})}\BibitemShut {NoStop}%
\bibitem [{\citenamefont {Mordacci}\ \emph {et~al.}(2024)\citenamefont
  {Mordacci}, \citenamefont {Ferrari},\ and\ \citenamefont
  {Amoretti}}]{Mordacci2024}%
  \BibitemOpen
  \bibfield  {author} {\bibinfo {author} {\bibfnamefont {M.}~\bibnamefont
  {Mordacci}}, \bibinfo {author} {\bibfnamefont {D.}~\bibnamefont {Ferrari}},\
  and\ \bibinfo {author} {\bibfnamefont {M.}~\bibnamefont {Amoretti}},\ }\href
  {https://arxiv.org/abs/2404.12741} {\bibinfo {title} {Multi-class quantum
  convolutional neural networks}} (\bibinfo {year} {2024}),\ \Eprint
  {https://arxiv.org/abs/2404.12741} {arXiv:2404.12741 [quant-ph]} \BibitemShut
  {NoStop}%
\bibitem [{\citenamefont {Cong}\ \emph {et~al.}(2019)\citenamefont {Cong},
  \citenamefont {Choi},\ and\ \citenamefont {Lukin}}]{Lukin2019}%
  \BibitemOpen
  \bibfield  {author} {\bibinfo {author} {\bibfnamefont {I.}~\bibnamefont
  {Cong}}, \bibinfo {author} {\bibfnamefont {S.}~\bibnamefont {Choi}},\ and\
  \bibinfo {author} {\bibfnamefont {M.~D.}\ \bibnamefont {Lukin}},\ }\bibfield
  {title} {\bibinfo {title} {Quantum convolutional neural networks},\ }\href
  {https://doi.org/10.1038/s41567-019-0648-8} {\bibfield  {journal} {\bibinfo
  {journal} {Nature Physics}\ }\textbf {\bibinfo {volume} {15}},\ \bibinfo
  {pages} {1273} (\bibinfo {year} {2019})}\BibitemShut {NoStop}%
\bibitem [{\citenamefont {MacCormack}\ \emph {et~al.}(2022)\citenamefont
  {MacCormack}, \citenamefont {Delaney}, \citenamefont {Galda}, \citenamefont
  {Aggarwal},\ and\ \citenamefont {Narang}}]{maccormack2022branching}%
  \BibitemOpen
  \bibfield  {author} {\bibinfo {author} {\bibfnamefont {I.}~\bibnamefont
  {MacCormack}}, \bibinfo {author} {\bibfnamefont {C.}~\bibnamefont {Delaney}},
  \bibinfo {author} {\bibfnamefont {A.}~\bibnamefont {Galda}}, \bibinfo
  {author} {\bibfnamefont {N.}~\bibnamefont {Aggarwal}},\ and\ \bibinfo
  {author} {\bibfnamefont {P.}~\bibnamefont {Narang}},\ }\bibfield  {title}
  {\bibinfo {title} {Branching quantum convolutional neural networks},\ }\href
  {https://doi.org/10.1103/PhysRevResearch.4.013117} {\bibfield  {journal}
  {\bibinfo  {journal} {Phys. Rev. Res.}\ }\textbf {\bibinfo {volume} {4}},\
  \bibinfo {pages} {013117} (\bibinfo {year} {2022})}\BibitemShut {NoStop}%
\bibitem [{\citenamefont {Herrmann}\ \emph {et~al.}(2022)\citenamefont
  {Herrmann}, \citenamefont {Llima}, \citenamefont {Remm}, \citenamefont
  {Zapletal}, \citenamefont {McMahon}, \citenamefont {Scarato}, \citenamefont
  {Swiadek}, \citenamefont {Andersen}, \citenamefont {Hellings}, \citenamefont
  {Krinner}, \citenamefont {Lacroix}, \citenamefont {Lazar}, \citenamefont
  {Kerschbaum}, \citenamefont {Zanuz}, \citenamefont {Norris}, \citenamefont
  {Hartmann}, \citenamefont {Wallraff},\ and\ \citenamefont
  {Eichler}}]{herrmann2022realizing}%
  \BibitemOpen
  \bibfield  {author} {\bibinfo {author} {\bibfnamefont {J.}~\bibnamefont
  {Herrmann}}, \bibinfo {author} {\bibfnamefont {S.~M.}\ \bibnamefont {Llima}},
  \bibinfo {author} {\bibfnamefont {A.}~\bibnamefont {Remm}}, \bibinfo {author}
  {\bibfnamefont {P.}~\bibnamefont {Zapletal}}, \bibinfo {author}
  {\bibfnamefont {N.~A.}\ \bibnamefont {McMahon}}, \bibinfo {author}
  {\bibfnamefont {C.}~\bibnamefont {Scarato}}, \bibinfo {author} {\bibfnamefont
  {F.}~\bibnamefont {Swiadek}}, \bibinfo {author} {\bibfnamefont {C.~K.}\
  \bibnamefont {Andersen}}, \bibinfo {author} {\bibfnamefont {C.}~\bibnamefont
  {Hellings}}, \bibinfo {author} {\bibfnamefont {S.}~\bibnamefont {Krinner}},
  \bibinfo {author} {\bibfnamefont {N.}~\bibnamefont {Lacroix}}, \bibinfo
  {author} {\bibfnamefont {S.}~\bibnamefont {Lazar}}, \bibinfo {author}
  {\bibfnamefont {M.}~\bibnamefont {Kerschbaum}}, \bibinfo {author}
  {\bibfnamefont {D.~C.}\ \bibnamefont {Zanuz}}, \bibinfo {author}
  {\bibfnamefont {G.~J.}\ \bibnamefont {Norris}}, \bibinfo {author}
  {\bibfnamefont {M.~J.}\ \bibnamefont {Hartmann}}, \bibinfo {author}
  {\bibfnamefont {A.}~\bibnamefont {Wallraff}},\ and\ \bibinfo {author}
  {\bibfnamefont {C.}~\bibnamefont {Eichler}},\ }\bibfield  {title} {\bibinfo
  {title} {Realizing quantum convolutional neural networks on a superconducting
  quantum processor to recognize quantum phases},\ }\href
  {https://doi.org/10.1038/s41467-022-31679-5} {\bibfield  {journal} {\bibinfo
  {journal} {Nature Communications}\ }\textbf {\bibinfo {volume} {13}},\
  \bibinfo {pages} {4144} (\bibinfo {year} {2022})}\BibitemShut {NoStop}%
\bibitem [{\citenamefont {Liu}\ \emph {et~al.}(2023)\citenamefont {Liu},
  \citenamefont {Smith}, \citenamefont {Knap},\ and\ \citenamefont
  {Pollmann}}]{liu2023model}%
  \BibitemOpen
  \bibfield  {author} {\bibinfo {author} {\bibfnamefont {Y.-J.}\ \bibnamefont
  {Liu}}, \bibinfo {author} {\bibfnamefont {A.}~\bibnamefont {Smith}}, \bibinfo
  {author} {\bibfnamefont {M.}~\bibnamefont {Knap}},\ and\ \bibinfo {author}
  {\bibfnamefont {F.}~\bibnamefont {Pollmann}},\ }\bibfield  {title} {\bibinfo
  {title} {Model-independent learning of quantum phases of matter with quantum
  convolutional neural networks},\ }\href
  {https://doi.org/10.1103/PhysRevLett.130.220603} {\bibfield  {journal}
  {\bibinfo  {journal} {Phys. Rev. Lett.}\ }\textbf {\bibinfo {volume} {130}},\
  \bibinfo {pages} {220603} (\bibinfo {year} {2023})}\BibitemShut {NoStop}%
\bibitem [{\citenamefont {Szo\l{}dra}\ \emph {et~al.}(2022)\citenamefont
  {Szo\l{}dra}, \citenamefont {Sierant}, \citenamefont {Lewenstein},\ and\
  \citenamefont {Zakrzewski}}]{Tomasz2022}%
  \BibitemOpen
  \bibfield  {author} {\bibinfo {author} {\bibfnamefont {T.}~\bibnamefont
  {Szo\l{}dra}}, \bibinfo {author} {\bibfnamefont {P.}~\bibnamefont {Sierant}},
  \bibinfo {author} {\bibfnamefont {M.}~\bibnamefont {Lewenstein}},\ and\
  \bibinfo {author} {\bibfnamefont {J.}~\bibnamefont {Zakrzewski}},\ }\bibfield
   {title} {\bibinfo {title} {Unsupervised detection of decoupled subspaces:
  Many-body scars and beyond},\ }\href
  {https://doi.org/10.1103/PhysRevB.105.224205} {\bibfield  {journal} {\bibinfo
   {journal} {Phys. Rev. B}\ }\textbf {\bibinfo {volume} {105}},\ \bibinfo
  {pages} {224205} (\bibinfo {year} {2022})}\BibitemShut {NoStop}%
\bibitem [{\citenamefont {Cao}\ \emph {et~al.}(2024)\citenamefont {Cao},
  \citenamefont {Angelakis},\ and\ \citenamefont {Leykam}}]{cao2024}%
  \BibitemOpen
  \bibfield  {author} {\bibinfo {author} {\bibfnamefont {H.}~\bibnamefont
  {Cao}}, \bibinfo {author} {\bibfnamefont {D.~G.}\ \bibnamefont {Angelakis}},\
  and\ \bibinfo {author} {\bibfnamefont {D.}~\bibnamefont {Leykam}},\
  }\bibfield  {title} {\bibinfo {title} {Unsupervised learning of quantum
  many-body scars using intrinsic dimension},\ }\href
  {https://doi.org/10.1088/2632-2153/ad4d3f} {\bibfield  {journal} {\bibinfo
  {journal} {Machine Learning: Science and Technology}\ }\textbf {\bibinfo
  {volume} {5}},\ \bibinfo {pages} {025049} (\bibinfo {year}
  {2024})}\BibitemShut {NoStop}%
\bibitem [{\citenamefont {Cenedese}\ \emph {et~al.}(2024)\citenamefont
  {Cenedese}, \citenamefont {Bondani}, \citenamefont {Andreanov}, \citenamefont
  {Carrega}, \citenamefont {Benenti},\ and\ \citenamefont
  {Rosa}}]{cenedese2024}%
  \BibitemOpen
  \bibfield  {author} {\bibinfo {author} {\bibfnamefont {G.}~\bibnamefont
  {Cenedese}}, \bibinfo {author} {\bibfnamefont {M.}~\bibnamefont {Bondani}},
  \bibinfo {author} {\bibfnamefont {A.}~\bibnamefont {Andreanov}}, \bibinfo
  {author} {\bibfnamefont {M.}~\bibnamefont {Carrega}}, \bibinfo {author}
  {\bibfnamefont {G.}~\bibnamefont {Benenti}},\ and\ \bibinfo {author}
  {\bibfnamefont {D.}~\bibnamefont {Rosa}},\ }\href
  {https://arxiv.org/abs/2401.09279} {\bibinfo {title} {Shallow quantum
  circuits are robust hunters for quantum many-body scars}} (\bibinfo {year}
  {2024}),\ \Eprint {https://arxiv.org/abs/2401.09279} {arXiv:2401.09279
  [quant-ph]} \BibitemShut {NoStop}%
\bibitem [{\citenamefont {Turner}\ \emph {et~al.}(2018)\citenamefont {Turner},
  \citenamefont {Michailidis}, \citenamefont {Abanin}, \citenamefont {Serbyn},\
  and\ \citenamefont {Papić}}]{Turner2018}%
  \BibitemOpen
  \bibfield  {author} {\bibinfo {author} {\bibfnamefont {C.~J.}\ \bibnamefont
  {Turner}}, \bibinfo {author} {\bibfnamefont {A.~A.}\ \bibnamefont
  {Michailidis}}, \bibinfo {author} {\bibfnamefont {D.~A.}\ \bibnamefont
  {Abanin}}, \bibinfo {author} {\bibfnamefont {M.}~\bibnamefont {Serbyn}},\
  and\ \bibinfo {author} {\bibfnamefont {Z.}~\bibnamefont {Papić}},\
  }\bibfield  {title} {\bibinfo {title} {Weak ergodicity breaking from quantum
  many-body scars},\ }\href {https://doi.org/10.1038/s41567-018-0137-5}
  {\bibfield  {journal} {\bibinfo  {journal} {Nature Physics}\ }\textbf
  {\bibinfo {volume} {14}},\ \bibinfo {pages} {745} (\bibinfo {year}
  {2018})}\BibitemShut {NoStop}%
\bibitem [{\citenamefont {Lin}\ and\ \citenamefont
  {Motrunich}(2019)}]{Lin2019}%
  \BibitemOpen
  \bibfield  {author} {\bibinfo {author} {\bibfnamefont {C.-J.}\ \bibnamefont
  {Lin}}\ and\ \bibinfo {author} {\bibfnamefont {O.~I.}\ \bibnamefont
  {Motrunich}},\ }\bibfield  {title} {\bibinfo {title} {Exact quantum many-body
  scar states in the rydberg-blockaded atom chain},\ }\href
  {https://doi.org/10.1103/PhysRevLett.122.173401} {\bibfield  {journal}
  {\bibinfo  {journal} {Phys. Rev. Lett.}\ }\textbf {\bibinfo {volume} {122}},\
  \bibinfo {pages} {173401} (\bibinfo {year} {2019})}\BibitemShut {NoStop}%
\bibitem [{\citenamefont {Zhang}\ \emph
  {et~al.}(2023{\natexlab{a}})\citenamefont {Zhang}, \citenamefont {Dong},
  \citenamefont {Gao}, \citenamefont {Zhao}, \citenamefont {Hao}, \citenamefont
  {Desaules}, \citenamefont {Guo}, \citenamefont {Chen}, \citenamefont {Deng},
  \citenamefont {Liu}, \citenamefont {Ren}, \citenamefont {Yao}, \citenamefont
  {Zhang}, \citenamefont {Xu}, \citenamefont {Wang}, \citenamefont {Jin},
  \citenamefont {Zhu}, \citenamefont {Zhang}, \citenamefont {Li}, \citenamefont
  {Song}, \citenamefont {Wang}, \citenamefont {Liu}, \citenamefont {Papić},
  \citenamefont {Ying}, \citenamefont {Wang},\ and\ \citenamefont
  {Lai}}]{ZhangPengfei2023}%
  \BibitemOpen
  \bibfield  {author} {\bibinfo {author} {\bibfnamefont {P.}~\bibnamefont
  {Zhang}}, \bibinfo {author} {\bibfnamefont {H.}~\bibnamefont {Dong}},
  \bibinfo {author} {\bibfnamefont {Y.}~\bibnamefont {Gao}}, \bibinfo {author}
  {\bibfnamefont {L.}~\bibnamefont {Zhao}}, \bibinfo {author} {\bibfnamefont
  {J.}~\bibnamefont {Hao}}, \bibinfo {author} {\bibfnamefont {J.-Y.}\
  \bibnamefont {Desaules}}, \bibinfo {author} {\bibfnamefont {Q.}~\bibnamefont
  {Guo}}, \bibinfo {author} {\bibfnamefont {J.}~\bibnamefont {Chen}}, \bibinfo
  {author} {\bibfnamefont {J.}~\bibnamefont {Deng}}, \bibinfo {author}
  {\bibfnamefont {B.}~\bibnamefont {Liu}}, \bibinfo {author} {\bibfnamefont
  {W.}~\bibnamefont {Ren}}, \bibinfo {author} {\bibfnamefont {Y.}~\bibnamefont
  {Yao}}, \bibinfo {author} {\bibfnamefont {X.}~\bibnamefont {Zhang}}, \bibinfo
  {author} {\bibfnamefont {S.}~\bibnamefont {Xu}}, \bibinfo {author}
  {\bibfnamefont {K.}~\bibnamefont {Wang}}, \bibinfo {author} {\bibfnamefont
  {F.}~\bibnamefont {Jin}}, \bibinfo {author} {\bibfnamefont {X.}~\bibnamefont
  {Zhu}}, \bibinfo {author} {\bibfnamefont {B.}~\bibnamefont {Zhang}}, \bibinfo
  {author} {\bibfnamefont {H.}~\bibnamefont {Li}}, \bibinfo {author}
  {\bibfnamefont {C.}~\bibnamefont {Song}}, \bibinfo {author} {\bibfnamefont
  {Z.}~\bibnamefont {Wang}}, \bibinfo {author} {\bibfnamefont {F.}~\bibnamefont
  {Liu}}, \bibinfo {author} {\bibfnamefont {Z.}~\bibnamefont {Papić}},
  \bibinfo {author} {\bibfnamefont {L.}~\bibnamefont {Ying}}, \bibinfo {author}
  {\bibfnamefont {H.}~\bibnamefont {Wang}},\ and\ \bibinfo {author}
  {\bibfnamefont {Y.-C.}\ \bibnamefont {Lai}},\ }\bibfield  {title} {\bibinfo
  {title} {Many-body hilbert space scarring on a superconducting processor},\
  }\href {https://doi.org/10.1038/s41567-022-01784-9} {\bibfield  {journal}
  {\bibinfo  {journal} {Nature Physics}\ }\textbf {\bibinfo {volume} {19}},\
  \bibinfo {pages} {120} (\bibinfo {year} {2023}{\natexlab{a}})}\BibitemShut
  {NoStop}%
\bibitem [{\citenamefont {Wang}\ \emph {et~al.}(2024)\citenamefont {Wang},
  \citenamefont {Zhou}, \citenamefont {Guo},\ and\ \citenamefont
  {Zhou}}]{WangJiaWei2024}%
  \BibitemOpen
  \bibfield  {author} {\bibinfo {author} {\bibfnamefont {J.-W.}\ \bibnamefont
  {Wang}}, \bibinfo {author} {\bibfnamefont {X.-F.}\ \bibnamefont {Zhou}},
  \bibinfo {author} {\bibfnamefont {G.-C.}\ \bibnamefont {Guo}},\ and\ \bibinfo
  {author} {\bibfnamefont {Z.-W.}\ \bibnamefont {Zhou}},\ }\bibfield  {title}
  {\bibinfo {title} {Quantum many-body scar models in one-dimensional spin
  chains},\ }\href {https://doi.org/10.1103/PhysRevB.109.125102} {\bibfield
  {journal} {\bibinfo  {journal} {Phys. Rev. B}\ }\textbf {\bibinfo {volume}
  {109}},\ \bibinfo {pages} {125102} (\bibinfo {year} {2024})}\BibitemShut
  {NoStop}%
\bibitem [{\citenamefont {Deger}\ and\ \citenamefont
  {Lazarides}(2024)}]{Deger2024}%
  \BibitemOpen
  \bibfield  {author} {\bibinfo {author} {\bibfnamefont {A.}~\bibnamefont
  {Deger}}\ and\ \bibinfo {author} {\bibfnamefont {A.}~\bibnamefont
  {Lazarides}},\ }\bibfield  {title} {\bibinfo {title} {Weak ergodicity
  breaking transition in a randomly constrained model},\ }\href
  {https://doi.org/10.1103/PhysRevB.109.L220301} {\bibfield  {journal}
  {\bibinfo  {journal} {Phys. Rev. B}\ }\textbf {\bibinfo {volume} {109}},\
  \bibinfo {pages} {L220301} (\bibinfo {year} {2024})}\BibitemShut {NoStop}%
\bibitem [{\citenamefont {Park}\ and\ \citenamefont {Lee}(2025)}]{HaRu2024}%
  \BibitemOpen
  \bibfield  {author} {\bibinfo {author} {\bibfnamefont {H.~K.}\ \bibnamefont
  {Park}}\ and\ \bibinfo {author} {\bibfnamefont {S.}~\bibnamefont {Lee}},\
  }\bibfield  {title} {\bibinfo {title} {Graph-theoretical proof of
  nonintegrability in quantum many-body systems: Application to the pxp
  model},\ }\href {https://doi.org/10.1103/PhysRevB.111.L081101} {\bibfield
  {journal} {\bibinfo  {journal} {Phys. Rev. B}\ }\textbf {\bibinfo {volume}
  {111}},\ \bibinfo {pages} {L081101} (\bibinfo {year} {2025})}\BibitemShut
  {NoStop}%
\bibitem [{\citenamefont {Buhrman}\ \emph {et~al.}(2022)\citenamefont
  {Buhrman}, \citenamefont {Linden}, \citenamefont {Mančinska}, \citenamefont
  {Montanaro},\ and\ \citenamefont {Ozols}}]{Harry2022}%
  \BibitemOpen
  \bibfield  {author} {\bibinfo {author} {\bibfnamefont {H.}~\bibnamefont
  {Buhrman}}, \bibinfo {author} {\bibfnamefont {N.}~\bibnamefont {Linden}},
  \bibinfo {author} {\bibfnamefont {L.}~\bibnamefont {Mančinska}}, \bibinfo
  {author} {\bibfnamefont {A.}~\bibnamefont {Montanaro}},\ and\ \bibinfo
  {author} {\bibfnamefont {M.}~\bibnamefont {Ozols}},\ }\href
  {https://arxiv.org/abs/2211.11729} {\bibinfo {title} {Quantum majority vote}}
  (\bibinfo {year} {2022}),\ \Eprint {https://arxiv.org/abs/2211.11729}
  {arXiv:2211.11729 [quant-ph]} \BibitemShut {NoStop}%
\bibitem [{\citenamefont {Gu}\ \emph {et~al.}(2024)\citenamefont {Gu},
  \citenamefont {Leone}, \citenamefont {Ghosh}, \citenamefont {Eisert},
  \citenamefont {Yelin},\ and\ \citenamefont {Quek}}]{Gu2024}%
  \BibitemOpen
  \bibfield  {author} {\bibinfo {author} {\bibfnamefont {A.}~\bibnamefont
  {Gu}}, \bibinfo {author} {\bibfnamefont {L.}~\bibnamefont {Leone}}, \bibinfo
  {author} {\bibfnamefont {S.}~\bibnamefont {Ghosh}}, \bibinfo {author}
  {\bibfnamefont {J.}~\bibnamefont {Eisert}}, \bibinfo {author} {\bibfnamefont
  {S.~F.}\ \bibnamefont {Yelin}},\ and\ \bibinfo {author} {\bibfnamefont
  {Y.}~\bibnamefont {Quek}},\ }\bibfield  {title} {\bibinfo {title}
  {Pseudomagic quantum states},\ }\href
  {https://doi.org/10.1103/PhysRevLett.132.210602} {\bibfield  {journal}
  {\bibinfo  {journal} {Phys. Rev. Lett.}\ }\textbf {\bibinfo {volume} {132}},\
  \bibinfo {pages} {210602} (\bibinfo {year} {2024})}\BibitemShut {NoStop}%
\bibitem [{\citenamefont {Mondragon-Shem}\ \emph {et~al.}(2021)\citenamefont
  {Mondragon-Shem}, \citenamefont {Vavilov},\ and\ \citenamefont
  {Martin}}]{Mondragon2021}%
  \BibitemOpen
  \bibfield  {author} {\bibinfo {author} {\bibfnamefont {I.}~\bibnamefont
  {Mondragon-Shem}}, \bibinfo {author} {\bibfnamefont {M.~G.}\ \bibnamefont
  {Vavilov}},\ and\ \bibinfo {author} {\bibfnamefont {I.}~\bibnamefont
  {Martin}},\ }\bibfield  {title} {\bibinfo {title} {Fate of quantum many-body
  scars in the presence of disorder},\ }\href
  {https://doi.org/10.1103/PRXQuantum.2.030349} {\bibfield  {journal} {\bibinfo
   {journal} {PRX Quantum}\ }\textbf {\bibinfo {volume} {2}},\ \bibinfo {pages}
  {030349} (\bibinfo {year} {2021})}\BibitemShut {NoStop}%
\bibitem [{\citenamefont {Liu}\ \emph {et~al.}(2019)\citenamefont {Liu},
  \citenamefont {Lundgren}, \citenamefont {Titum}, \citenamefont {Pagano},
  \citenamefont {Zhang}, \citenamefont {Monroe},\ and\ \citenamefont
  {Gorshkov}}]{PhysRevLett.122.150601}%
  \BibitemOpen
  \bibfield  {author} {\bibinfo {author} {\bibfnamefont {F.}~\bibnamefont
  {Liu}}, \bibinfo {author} {\bibfnamefont {R.}~\bibnamefont {Lundgren}},
  \bibinfo {author} {\bibfnamefont {P.}~\bibnamefont {Titum}}, \bibinfo
  {author} {\bibfnamefont {G.}~\bibnamefont {Pagano}}, \bibinfo {author}
  {\bibfnamefont {J.}~\bibnamefont {Zhang}}, \bibinfo {author} {\bibfnamefont
  {C.}~\bibnamefont {Monroe}},\ and\ \bibinfo {author} {\bibfnamefont {A.~V.}\
  \bibnamefont {Gorshkov}},\ }\bibfield  {title} {\bibinfo {title} {Confined
  quasiparticle dynamics in long-range interacting quantum spin chains},\
  }\href {https://doi.org/10.1103/PhysRevLett.122.150601} {\bibfield  {journal}
  {\bibinfo  {journal} {Phys. Rev. Lett.}\ }\textbf {\bibinfo {volume} {122}},\
  \bibinfo {pages} {150601} (\bibinfo {year} {2019})}\BibitemShut {NoStop}%
\bibitem [{\citenamefont {Wang}\ \emph
  {et~al.}(2021{\natexlab{b}})\citenamefont {Wang}, \citenamefont {Wang},\ and\
  \citenamefont {Wu}}]{Wang2021D}%
  \BibitemOpen
  \bibfield  {author} {\bibinfo {author} {\bibfnamefont {Z.}~\bibnamefont
  {Wang}}, \bibinfo {author} {\bibfnamefont {Y.}~\bibnamefont {Wang}},\ and\
  \bibinfo {author} {\bibfnamefont {B.}~\bibnamefont {Wu}},\ }\bibfield
  {title} {\bibinfo {title} {Quantum chaos and physical distance between
  quantum states},\ }\href {https://doi.org/10.1103/PhysRevE.103.042209}
  {\bibfield  {journal} {\bibinfo  {journal} {Phys. Rev. E}\ }\textbf {\bibinfo
  {volume} {103}},\ \bibinfo {pages} {042209} (\bibinfo {year}
  {2021}{\natexlab{b}})}\BibitemShut {NoStop}%
\bibitem [{\citenamefont {Birnkammer}\ \emph {et~al.}(2022)\citenamefont
  {Birnkammer}, \citenamefont {Bastianello},\ and\ \citenamefont
  {Knap}}]{Birnkammer2022}%
  \BibitemOpen
  \bibfield  {author} {\bibinfo {author} {\bibfnamefont {S.}~\bibnamefont
  {Birnkammer}}, \bibinfo {author} {\bibfnamefont {A.}~\bibnamefont
  {Bastianello}},\ and\ \bibinfo {author} {\bibfnamefont {M.}~\bibnamefont
  {Knap}},\ }\bibfield  {title} {\bibinfo {title} {Prethermalization in
  one-dimensional quantum many-body systems with confinement},\ }\href
  {https://doi.org/10.1038/s41467-022-35301-6} {\bibfield  {journal} {\bibinfo
  {journal} {Nature Communications}\ }\textbf {\bibinfo {volume} {13}},\
  \bibinfo {pages} {7663} (\bibinfo {year} {2022})}\BibitemShut {NoStop}%
\bibitem [{\citenamefont {Turner}\ \emph {et~al.}(2021)\citenamefont {Turner},
  \citenamefont {Desaules}, \citenamefont {Bull},\ and\ \citenamefont
  {Papi\ifmmode~\acute{c}\else \'{c}\fi{}}}]{Turner2021}%
  \BibitemOpen
  \bibfield  {author} {\bibinfo {author} {\bibfnamefont {C.~J.}\ \bibnamefont
  {Turner}}, \bibinfo {author} {\bibfnamefont {J.-Y.}\ \bibnamefont
  {Desaules}}, \bibinfo {author} {\bibfnamefont {K.}~\bibnamefont {Bull}},\
  and\ \bibinfo {author} {\bibfnamefont {Z.}~\bibnamefont
  {Papi\ifmmode~\acute{c}\else \'{c}\fi{}}},\ }\bibfield  {title} {\bibinfo
  {title} {Correspondence principle for many-body scars in ultracold rydberg
  atoms},\ }\href {https://doi.org/10.1103/PhysRevX.11.021021} {\bibfield
  {journal} {\bibinfo  {journal} {Phys. Rev. X}\ }\textbf {\bibinfo {volume}
  {11}},\ \bibinfo {pages} {021021} (\bibinfo {year} {2021})}\BibitemShut
  {NoStop}%
\bibitem [{\citenamefont {Choi}\ \emph {et~al.}(2019)\citenamefont {Choi},
  \citenamefont {Turner}, \citenamefont {Pichler}, \citenamefont {Ho},
  \citenamefont {Michailidis}, \citenamefont {Papi\ifmmode~\acute{c}\else
  \'{c}\fi{}}, \citenamefont {Serbyn}, \citenamefont {Lukin},\ and\
  \citenamefont {Abanin}}]{Choi2019}%
  \BibitemOpen
  \bibfield  {author} {\bibinfo {author} {\bibfnamefont {S.}~\bibnamefont
  {Choi}}, \bibinfo {author} {\bibfnamefont {C.~J.}\ \bibnamefont {Turner}},
  \bibinfo {author} {\bibfnamefont {H.}~\bibnamefont {Pichler}}, \bibinfo
  {author} {\bibfnamefont {W.~W.}\ \bibnamefont {Ho}}, \bibinfo {author}
  {\bibfnamefont {A.~A.}\ \bibnamefont {Michailidis}}, \bibinfo {author}
  {\bibfnamefont {Z.}~\bibnamefont {Papi\ifmmode~\acute{c}\else \'{c}\fi{}}},
  \bibinfo {author} {\bibfnamefont {M.}~\bibnamefont {Serbyn}}, \bibinfo
  {author} {\bibfnamefont {M.~D.}\ \bibnamefont {Lukin}},\ and\ \bibinfo
  {author} {\bibfnamefont {D.~A.}\ \bibnamefont {Abanin}},\ }\bibfield  {title}
  {\bibinfo {title} {Emergent su(2) dynamics and perfect quantum many-body
  scars},\ }\href {https://doi.org/10.1103/PhysRevLett.122.220603} {\bibfield
  {journal} {\bibinfo  {journal} {Phys. Rev. Lett.}\ }\textbf {\bibinfo
  {volume} {122}},\ \bibinfo {pages} {220603} (\bibinfo {year}
  {2019})}\BibitemShut {NoStop}%
\bibitem [{\citenamefont {Khemani}\ \emph {et~al.}(2019)\citenamefont
  {Khemani}, \citenamefont {Laumann},\ and\ \citenamefont
  {Chandran}}]{Khemani2019}%
  \BibitemOpen
  \bibfield  {author} {\bibinfo {author} {\bibfnamefont {V.}~\bibnamefont
  {Khemani}}, \bibinfo {author} {\bibfnamefont {C.~R.}\ \bibnamefont
  {Laumann}},\ and\ \bibinfo {author} {\bibfnamefont {A.}~\bibnamefont
  {Chandran}},\ }\bibfield  {title} {\bibinfo {title} {Signatures of
  integrability in the dynamics of rydberg-blockaded chains},\ }\href
  {https://doi.org/10.1103/PhysRevB.99.161101} {\bibfield  {journal} {\bibinfo
  {journal} {Phys. Rev. B}\ }\textbf {\bibinfo {volume} {99}},\ \bibinfo
  {pages} {161101} (\bibinfo {year} {2019})}\BibitemShut {NoStop}%
\bibitem [{\citenamefont {Desaules}\ \emph {et~al.}(2024)\citenamefont
  {Desaules}, \citenamefont {Su}, \citenamefont {McCulloch}, \citenamefont
  {Yang}, \citenamefont {Papi{\'{c}}},\ and\ \citenamefont
  {Halimeh}}]{Desaules2024}%
  \BibitemOpen
  \bibfield  {author} {\bibinfo {author} {\bibfnamefont {J.-Y.}\ \bibnamefont
  {Desaules}}, \bibinfo {author} {\bibfnamefont {G.-X.}\ \bibnamefont {Su}},
  \bibinfo {author} {\bibfnamefont {I.~P.}\ \bibnamefont {McCulloch}}, \bibinfo
  {author} {\bibfnamefont {B.}~\bibnamefont {Yang}}, \bibinfo {author}
  {\bibfnamefont {Z.}~\bibnamefont {Papi{\'{c}}}},\ and\ \bibinfo {author}
  {\bibfnamefont {J.~C.}\ \bibnamefont {Halimeh}},\ }\bibfield  {title}
  {\bibinfo {title} {Ergodicity {B}reaking {U}nder {C}onfinement in
  {C}old-{A}tom {Q}uantum {S}imulators},\ }\href
  {https://doi.org/10.22331/q-2024-02-29-1274} {\bibfield  {journal} {\bibinfo
  {journal} {{Quantum}}\ }\textbf {\bibinfo {volume} {8}},\ \bibinfo {pages}
  {1274} (\bibinfo {year} {2024})}\BibitemShut {NoStop}%
\bibitem [{\citenamefont {Han}\ \emph {et~al.}(2023)\citenamefont {Han},
  \citenamefont {Wang},\ and\ \citenamefont {Lai}}]{han2023meta}%
  \BibitemOpen
  \bibfield  {author} {\bibinfo {author} {\bibfnamefont {C.-D.}\ \bibnamefont
  {Han}}, \bibinfo {author} {\bibfnamefont {C.-Z.}\ \bibnamefont {Wang}},\ and\
  \bibinfo {author} {\bibfnamefont {Y.-C.}\ \bibnamefont {Lai}},\ }\bibfield
  {title} {\bibinfo {title} {Meta-machine-learning-based quantum scar
  detector},\ }\href {https://doi.org/10.1103/PhysRevApplied.19.064042}
  {\bibfield  {journal} {\bibinfo  {journal} {Phys. Rev. Appl.}\ }\textbf
  {\bibinfo {volume} {19}},\ \bibinfo {pages} {064042} (\bibinfo {year}
  {2023})}\BibitemShut {NoStop}%
\bibitem [{\citenamefont {Bermejo}\ \emph {et~al.}(2024)\citenamefont
  {Bermejo}, \citenamefont {Braccia}, \citenamefont {Rudolph}, \citenamefont
  {Holmes}, \citenamefont {Cincio},\ and\ \citenamefont
  {Cerezo}}]{bermejo2024quantum}%
  \BibitemOpen
  \bibfield  {author} {\bibinfo {author} {\bibfnamefont {P.}~\bibnamefont
  {Bermejo}}, \bibinfo {author} {\bibfnamefont {P.}~\bibnamefont {Braccia}},
  \bibinfo {author} {\bibfnamefont {M.~S.}\ \bibnamefont {Rudolph}}, \bibinfo
  {author} {\bibfnamefont {Z.}~\bibnamefont {Holmes}}, \bibinfo {author}
  {\bibfnamefont {L.}~\bibnamefont {Cincio}},\ and\ \bibinfo {author}
  {\bibfnamefont {M.}~\bibnamefont {Cerezo}},\ }\href
  {https://arxiv.org/abs/2408.12739} {\bibinfo {title} {Quantum convolutional
  neural networks are (effectively) classically simulable}} (\bibinfo {year}
  {2024}),\ \Eprint {https://arxiv.org/abs/2408.12739} {arXiv:2408.12739
  [quant-ph]} \BibitemShut {NoStop}%
\bibitem [{\citenamefont {Tilly}\ \emph {et~al.}(2022)\citenamefont {Tilly},
  \citenamefont {Chen}, \citenamefont {Cao}, \citenamefont {Picozzi},
  \citenamefont {Setia}, \citenamefont {Li}, \citenamefont {Grant},
  \citenamefont {Wossnig}, \citenamefont {Rungger}, \citenamefont {Booth} \emph
  {et~al.}}]{tilly2022variational}%
  \BibitemOpen
  \bibfield  {author} {\bibinfo {author} {\bibfnamefont {J.}~\bibnamefont
  {Tilly}}, \bibinfo {author} {\bibfnamefont {H.}~\bibnamefont {Chen}},
  \bibinfo {author} {\bibfnamefont {S.}~\bibnamefont {Cao}}, \bibinfo {author}
  {\bibfnamefont {D.}~\bibnamefont {Picozzi}}, \bibinfo {author} {\bibfnamefont
  {K.}~\bibnamefont {Setia}}, \bibinfo {author} {\bibfnamefont
  {Y.}~\bibnamefont {Li}}, \bibinfo {author} {\bibfnamefont {E.}~\bibnamefont
  {Grant}}, \bibinfo {author} {\bibfnamefont {L.}~\bibnamefont {Wossnig}},
  \bibinfo {author} {\bibfnamefont {I.}~\bibnamefont {Rungger}}, \bibinfo
  {author} {\bibfnamefont {G.~H.}\ \bibnamefont {Booth}}, \emph {et~al.},\
  }\bibfield  {title} {\bibinfo {title} {The variational quantum eigensolver: a
  review of methods and best practices},\ }\href@noop {} {\bibfield  {journal}
  {\bibinfo  {journal} {Physics Reports}\ }\textbf {\bibinfo {volume} {986}},\
  \bibinfo {pages} {1} (\bibinfo {year} {2022})}\BibitemShut {NoStop}%
\bibitem [{\citenamefont {Ho}\ \emph {et~al.}(2019)\citenamefont {Ho},
  \citenamefont {Choi}, \citenamefont {Pichler},\ and\ \citenamefont
  {Lukin}}]{Ho2019}%
  \BibitemOpen
  \bibfield  {author} {\bibinfo {author} {\bibfnamefont {W.~W.}\ \bibnamefont
  {Ho}}, \bibinfo {author} {\bibfnamefont {S.}~\bibnamefont {Choi}}, \bibinfo
  {author} {\bibfnamefont {H.}~\bibnamefont {Pichler}},\ and\ \bibinfo {author}
  {\bibfnamefont {M.~D.}\ \bibnamefont {Lukin}},\ }\bibfield  {title} {\bibinfo
  {title} {Periodic orbits, entanglement, and quantum many-body scars in
  constrained models: Matrix product state approach},\ }\href
  {https://doi.org/10.1103/PhysRevLett.122.040603} {\bibfield  {journal}
  {\bibinfo  {journal} {Phys. Rev. Lett.}\ }\textbf {\bibinfo {volume} {122}},\
  \bibinfo {pages} {040603} (\bibinfo {year} {2019})}\BibitemShut {NoStop}%
\bibitem [{\citenamefont {Zhang}\ \emph
  {et~al.}(2023{\natexlab{b}})\citenamefont {Zhang}, \citenamefont {Yuan},
  \citenamefont {Iadecola}, \citenamefont {Xu},\ and\ \citenamefont
  {Deng}}]{Zhang2023}%
  \BibitemOpen
  \bibfield  {author} {\bibinfo {author} {\bibfnamefont {S.-Y.}\ \bibnamefont
  {Zhang}}, \bibinfo {author} {\bibfnamefont {D.}~\bibnamefont {Yuan}},
  \bibinfo {author} {\bibfnamefont {T.}~\bibnamefont {Iadecola}}, \bibinfo
  {author} {\bibfnamefont {S.}~\bibnamefont {Xu}},\ and\ \bibinfo {author}
  {\bibfnamefont {D.-L.}\ \bibnamefont {Deng}},\ }\bibfield  {title} {\bibinfo
  {title} {Extracting quantum many-body scarred eigenstates with matrix product
  states},\ }\href {https://doi.org/10.1103/PhysRevLett.131.020402} {\bibfield
  {journal} {\bibinfo  {journal} {Phys. Rev. Lett.}\ }\textbf {\bibinfo
  {volume} {131}},\ \bibinfo {pages} {020402} (\bibinfo {year}
  {2023}{\natexlab{b}})}\BibitemShut {NoStop}%
\bibitem [{\citenamefont {Malz}\ \emph {et~al.}(2024)\citenamefont {Malz},
  \citenamefont {Styliaris}, \citenamefont {Wei},\ and\ \citenamefont
  {Cirac}}]{Malz2024}%
  \BibitemOpen
  \bibfield  {author} {\bibinfo {author} {\bibfnamefont {D.}~\bibnamefont
  {Malz}}, \bibinfo {author} {\bibfnamefont {G.}~\bibnamefont {Styliaris}},
  \bibinfo {author} {\bibfnamefont {Z.-Y.}\ \bibnamefont {Wei}},\ and\ \bibinfo
  {author} {\bibfnamefont {J.~I.}\ \bibnamefont {Cirac}},\ }\bibfield  {title}
  {\bibinfo {title} {Preparation of matrix product states with log-depth
  quantum circuits},\ }\href {https://doi.org/10.1103/PhysRevLett.132.040404}
  {\bibfield  {journal} {\bibinfo  {journal} {Phys. Rev. Lett.}\ }\textbf
  {\bibinfo {volume} {132}},\ \bibinfo {pages} {040404} (\bibinfo {year}
  {2024})}\BibitemShut {NoStop}%
\bibitem [{\citenamefont {Caro}\ \emph {et~al.}(2022)\citenamefont {Caro},
  \citenamefont {Huang}, \citenamefont {Cerezo}, \citenamefont {Sharma},
  \citenamefont {Sornborger}, \citenamefont {Cincio},\ and\ \citenamefont
  {Coles}}]{Caro2022}%
  \BibitemOpen
  \bibfield  {author} {\bibinfo {author} {\bibfnamefont {M.~C.}\ \bibnamefont
  {Caro}}, \bibinfo {author} {\bibfnamefont {H.-Y.}\ \bibnamefont {Huang}},
  \bibinfo {author} {\bibfnamefont {M.}~\bibnamefont {Cerezo}}, \bibinfo
  {author} {\bibfnamefont {K.}~\bibnamefont {Sharma}}, \bibinfo {author}
  {\bibfnamefont {A.}~\bibnamefont {Sornborger}}, \bibinfo {author}
  {\bibfnamefont {L.}~\bibnamefont {Cincio}},\ and\ \bibinfo {author}
  {\bibfnamefont {P.~J.}\ \bibnamefont {Coles}},\ }\bibfield  {title} {\bibinfo
  {title} {Generalization in quantum machine learning from few training data},\
  }\href {https://doi.org/10.1038/s41467-022-32550-3} {\bibfield  {journal}
  {\bibinfo  {journal} {Nature Communications}\ }\textbf {\bibinfo {volume}
  {13}},\ \bibinfo {pages} {4919} (\bibinfo {year} {2022})}\BibitemShut
  {NoStop}%
\bibitem [{\citenamefont {Cerezo}\ \emph {et~al.}(2021)\citenamefont {Cerezo},
  \citenamefont {Sone}, \citenamefont {Volkoff}, \citenamefont {Cincio},\ and\
  \citenamefont {Coles}}]{Cerezo2021}%
  \BibitemOpen
  \bibfield  {author} {\bibinfo {author} {\bibfnamefont {M.}~\bibnamefont
  {Cerezo}}, \bibinfo {author} {\bibfnamefont {A.}~\bibnamefont {Sone}},
  \bibinfo {author} {\bibfnamefont {T.}~\bibnamefont {Volkoff}}, \bibinfo
  {author} {\bibfnamefont {L.}~\bibnamefont {Cincio}},\ and\ \bibinfo {author}
  {\bibfnamefont {P.~J.}\ \bibnamefont {Coles}},\ }\bibfield  {title} {\bibinfo
  {title} {Cost function dependent barren plateaus in shallow parametrized
  quantum circuits},\ }\href {https://doi.org/10.1038/s41467-021-21728-w}
  {\bibfield  {journal} {\bibinfo  {journal} {Nature Communications}\ }\textbf
  {\bibinfo {volume} {12}},\ \bibinfo {pages} {1791} (\bibinfo {year}
  {2021})}\BibitemShut {NoStop}%
\bibitem [{\citenamefont {Gustafson}\ \emph {et~al.}(2023)\citenamefont
  {Gustafson}, \citenamefont {Li}, \citenamefont {Khan}, \citenamefont {Kim},
  \citenamefont {Kurkcuoglu}, \citenamefont {Alam}, \citenamefont {Orth},
  \citenamefont {Rahmani},\ and\ \citenamefont {Iadecola}}]{Gustafson2023}%
  \BibitemOpen
  \bibfield  {author} {\bibinfo {author} {\bibfnamefont {E.~J.}\ \bibnamefont
  {Gustafson}}, \bibinfo {author} {\bibfnamefont {A.~C.~Y.}\ \bibnamefont
  {Li}}, \bibinfo {author} {\bibfnamefont {A.}~\bibnamefont {Khan}}, \bibinfo
  {author} {\bibfnamefont {J.}~\bibnamefont {Kim}}, \bibinfo {author}
  {\bibfnamefont {D.~M.}\ \bibnamefont {Kurkcuoglu}}, \bibinfo {author}
  {\bibfnamefont {M.~S.}\ \bibnamefont {Alam}}, \bibinfo {author}
  {\bibfnamefont {P.~P.}\ \bibnamefont {Orth}}, \bibinfo {author}
  {\bibfnamefont {A.}~\bibnamefont {Rahmani}},\ and\ \bibinfo {author}
  {\bibfnamefont {T.}~\bibnamefont {Iadecola}},\ }\bibfield  {title} {\bibinfo
  {title} {Preparing quantum many-body scar states on quantum computers},\
  }\href {https://doi.org/10.22331/q-2023-11-07-1171} {\bibfield  {journal}
  {\bibinfo  {journal} {{Quantum}}\ }\textbf {\bibinfo {volume} {7}},\ \bibinfo
  {pages} {1171} (\bibinfo {year} {2023})}\BibitemShut {NoStop}%
\bibitem [{\citenamefont {Qin}\ \emph {et~al.}(2022)\citenamefont {Qin},
  \citenamefont {Xu},\ and\ \citenamefont {Li}}]{Qin2022}%
  \BibitemOpen
  \bibfield  {author} {\bibinfo {author} {\bibfnamefont {D.}~\bibnamefont
  {Qin}}, \bibinfo {author} {\bibfnamefont {X.}~\bibnamefont {Xu}},\ and\
  \bibinfo {author} {\bibfnamefont {Y.}~\bibnamefont {Li}},\ }\bibfield
  {title} {\bibinfo {title} {An overview of quantum error mitigation
  formulas},\ }\href {https://doi.org/10.1088/1674-1056/ac7b1e} {\bibfield
  {journal} {\bibinfo  {journal} {Chinese Physics B}\ }\textbf {\bibinfo
  {volume} {31}},\ \bibinfo {pages} {090306} (\bibinfo {year}
  {2022})}\BibitemShut {NoStop}%
\bibitem [{\citenamefont {Daniel}\ \emph {et~al.}(2023)\citenamefont {Daniel},
  \citenamefont {Hallam}, \citenamefont {Desaules}, \citenamefont {Hudomal},
  \citenamefont {Su}, \citenamefont {Halimeh},\ and\ \citenamefont
  {Papi\ifmmode~\acute{c}\else \'{c}\fi{}}}]{Daniel2023}%
  \BibitemOpen
  \bibfield  {author} {\bibinfo {author} {\bibfnamefont {A.}~\bibnamefont
  {Daniel}}, \bibinfo {author} {\bibfnamefont {A.}~\bibnamefont {Hallam}},
  \bibinfo {author} {\bibfnamefont {J.-Y.}\ \bibnamefont {Desaules}}, \bibinfo
  {author} {\bibfnamefont {A.}~\bibnamefont {Hudomal}}, \bibinfo {author}
  {\bibfnamefont {G.-X.}\ \bibnamefont {Su}}, \bibinfo {author} {\bibfnamefont
  {J.~C.}\ \bibnamefont {Halimeh}},\ and\ \bibinfo {author} {\bibfnamefont
  {Z.}~\bibnamefont {Papi\ifmmode~\acute{c}\else \'{c}\fi{}}},\ }\bibfield
  {title} {\bibinfo {title} {Bridging quantum criticality via many-body
  scarring},\ }\href {https://doi.org/10.1103/PhysRevB.107.235108} {\bibfield
  {journal} {\bibinfo  {journal} {Phys. Rev. B}\ }\textbf {\bibinfo {volume}
  {107}},\ \bibinfo {pages} {235108} (\bibinfo {year} {2023})}\BibitemShut
  {NoStop}%
\bibitem [{\citenamefont {Su}\ \emph {et~al.}(2023)\citenamefont {Su},
  \citenamefont {Sun}, \citenamefont {Hudomal}, \citenamefont {Desaules},
  \citenamefont {Zhou}, \citenamefont {Yang}, \citenamefont {Halimeh},
  \citenamefont {Yuan}, \citenamefont {Papi\ifmmode~\acute{c}\else
  \'{c}\fi{}},\ and\ \citenamefont {Pan}}]{Su2023}%
  \BibitemOpen
  \bibfield  {author} {\bibinfo {author} {\bibfnamefont {G.-X.}\ \bibnamefont
  {Su}}, \bibinfo {author} {\bibfnamefont {H.}~\bibnamefont {Sun}}, \bibinfo
  {author} {\bibfnamefont {A.}~\bibnamefont {Hudomal}}, \bibinfo {author}
  {\bibfnamefont {J.-Y.}\ \bibnamefont {Desaules}}, \bibinfo {author}
  {\bibfnamefont {Z.-Y.}\ \bibnamefont {Zhou}}, \bibinfo {author}
  {\bibfnamefont {B.}~\bibnamefont {Yang}}, \bibinfo {author} {\bibfnamefont
  {J.~C.}\ \bibnamefont {Halimeh}}, \bibinfo {author} {\bibfnamefont {Z.-S.}\
  \bibnamefont {Yuan}}, \bibinfo {author} {\bibfnamefont {Z.}~\bibnamefont
  {Papi\ifmmode~\acute{c}\else \'{c}\fi{}}},\ and\ \bibinfo {author}
  {\bibfnamefont {J.-W.}\ \bibnamefont {Pan}},\ }\bibfield  {title} {\bibinfo
  {title} {Observation of many-body scarring in a bose-hubbard quantum
  simulator},\ }\href {https://doi.org/10.1103/PhysRevResearch.5.023010}
  {\bibfield  {journal} {\bibinfo  {journal} {Phys. Rev. Res.}\ }\textbf
  {\bibinfo {volume} {5}},\ \bibinfo {pages} {023010} (\bibinfo {year}
  {2023})}\BibitemShut {NoStop}%
\end{thebibliography}

%

\appendix

\section{Details on the QCNN \label{App:QCNN}}

\begin{figure}[b]
\begin{center}
\includegraphics[clip = true, width =\columnwidth]{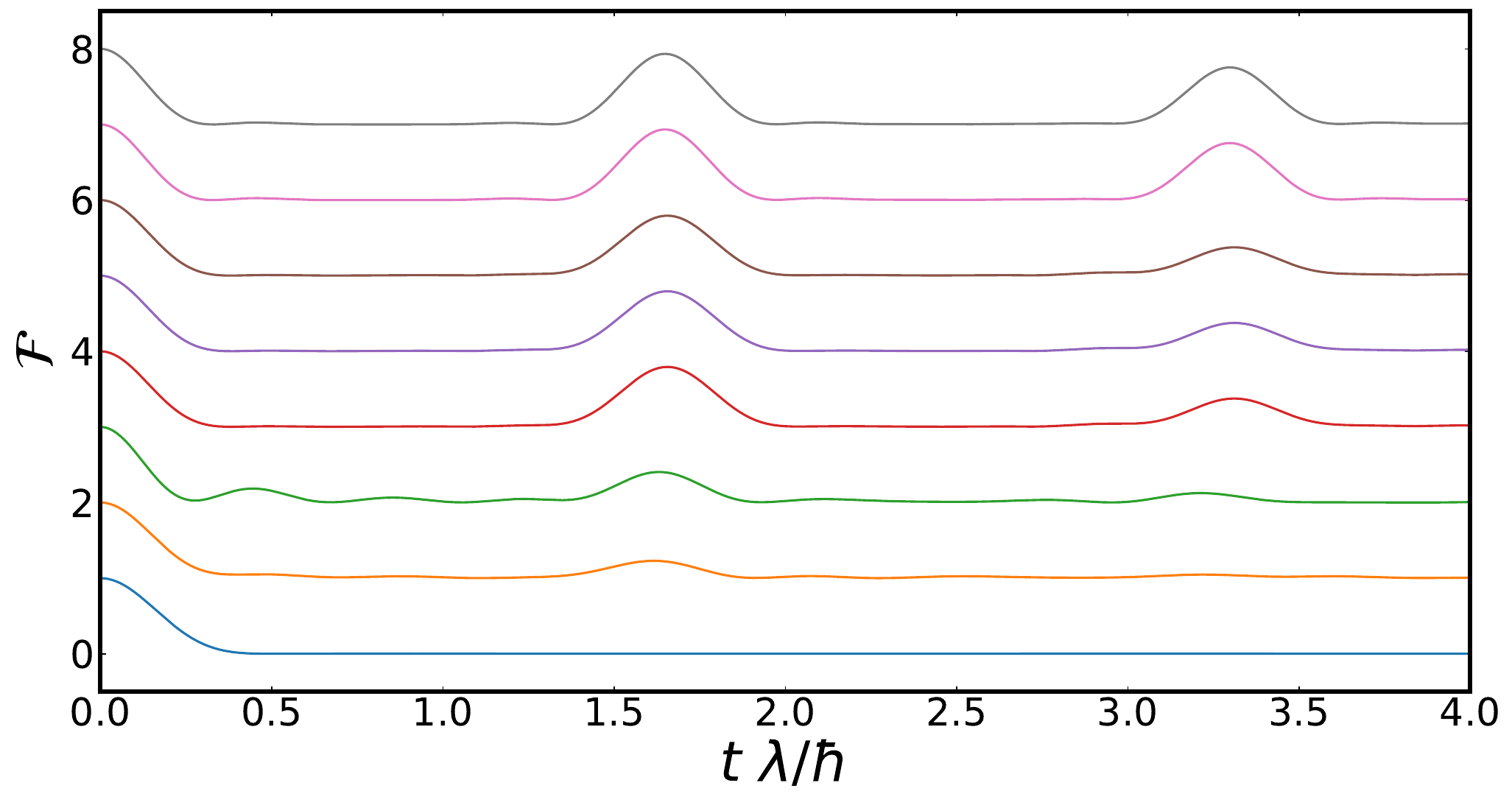}
\caption{\label{fig:MultiEvo} (color online) Revival dynamics for various cases corresponding to the points in Fig. 3. The curves are vertically offset for clarity. The curves are vertically offset for clarity. From bottom to top, the number of convolutional layers before the pooling layers increases from 0 to 7.}
\end{center}
\end{figure}

\begin{figure*}[tb]
\begin{center}
\includegraphics[clip = true, width =2\columnwidth]{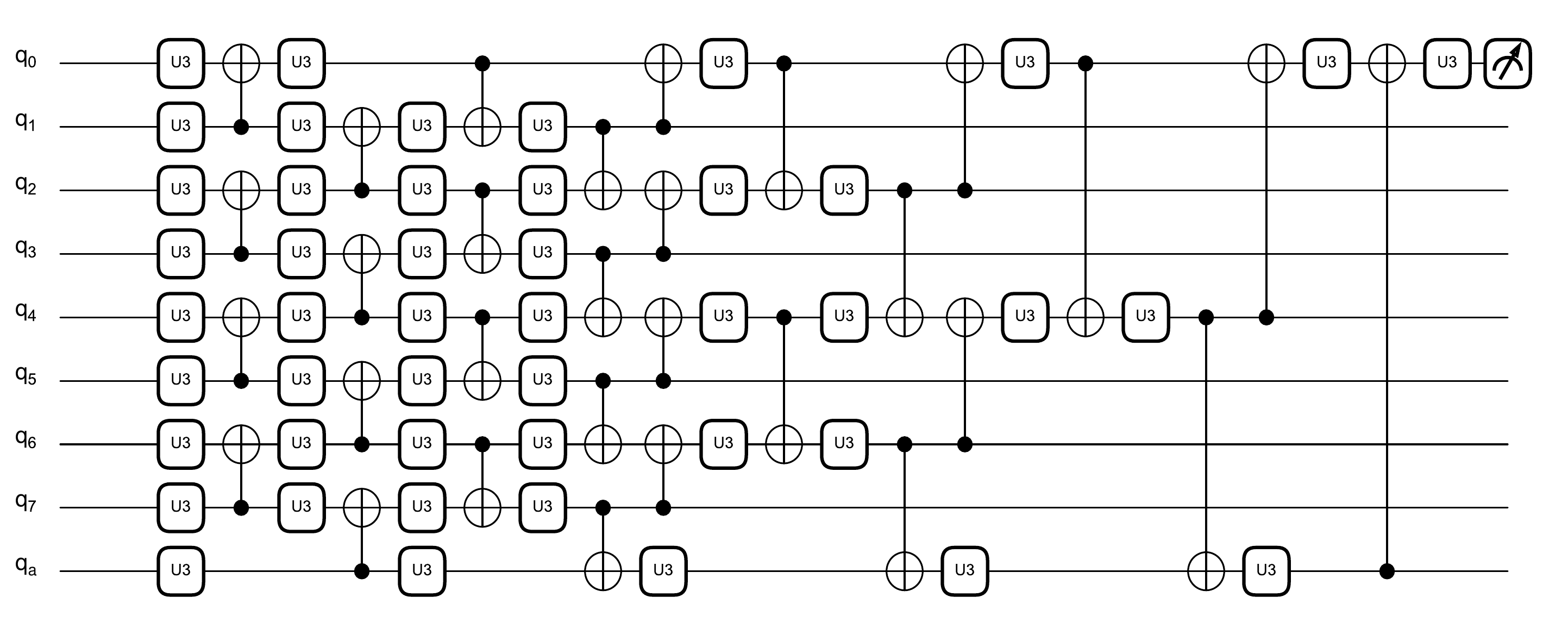}
\caption{\label{fig:IBMQCNN} Circuit diagram of the QCNN used in the experiment. U3 represent single-qubit gates with three rotation angles. The single-qubit gates in the first two columns have independent parameters, while the gates in subsequent columns share parameters with those in the same column. }
\end{center}
\end{figure*}

In this section, we provide a detailed description of the QCNN used in our work. The QCNN is implemented as a parameterized quantum circuit, incorporating general single-qubit rotations and two-qubit rotation gates. In our simulations, we employed rotation-XX, YY, and ZZ gates on nearest-neighbor qubits \cite{Caro2022}. To mitigate potential trainability issues, the pre-processing circuit is designed with a fixed depth that remains constant regardless of the number of qubits.

In classical convolutional neural networks, zero padding is used to control the spatial dimensions of the output feature map by adding extra pixels to the edges of the input image. Analogously, in QCNNs, we introduce an ancilla qubit initialized in the state $|0\rangle$ manage the size of the Hilbert space.

The QCNN employs a single output qubit to perform classification. The ancilla qubit becomes necessary when the mapping is not uniformly distributed within the Hilbert space.
Ideally, we aim for the unitary operation to satisfy the following classification equation:
\begin{eqnarray}
U| A_i \rangle &=& |a_i\rangle\otimes|0\rangle ~, \\
U| B_j \rangle &=& |b_j\rangle\otimes|1\rangle ~,
\end{eqnarray}
where $| A_i \rangle$ and $| B_j \rangle$ are $n$-qubit states that span the entire Hilbert space (i.e., by varying all $i$, $j$). Meanwhile, $| a_i \rangle$ and $| b_j \rangle$ represent $(n-1)$-qubit states. Note that the right-hand side of the equations occupies less than half of the Hilbert space. If the left-hand side exceeds half of the Hilbert space, no unitary operation $U$ can achieve this mapping, as unitary operations do not alter the size of the Hilbert space. To address this, we introduce an ancilla qubit initialized to the state $\ket{0}$, a common technique in quantum computing \cite{Tomasz2022}. This modification adjusts the equations as follows:
\begin{eqnarray}
U| A_i \rangle\otimes|0\rangle &=& |\alpha_i\rangle\otimes|0\rangle ~, \\
U| B_j \rangle\otimes|0\rangle &=& |\beta_j\rangle\otimes|1\rangle ~,
\end{eqnarray}
where $| \alpha_i \rangle$ and $| \beta_j \rangle$ are $n$-qubits states. With this adjustment, both the left-hand side and right-hand side of the equations now occupy the same size of Hilbert space.

The performance of the circuit is also influenced by the number of trainable parameters. As the number of parameters increases, the criteria for integrability become more stringent. This leads to more pronounced revivals, as illustrated in Fig. \ref{fig:MultiEvo}.

\section{Details of experiments \label{App:Exp}}

We utilized the IBM hanoi quantum device for our experiments. Its basic two-qubit gate, the CNOT gate, has an error probability of approximately 1\%. To align the QCNN architecture with the hardware capabilities, we simplified its structure as depicted in Fig.~\ref{fig:IBMQCNN}. Additionally, we incorporated general layers to address finite-size effects and enhance robustness against noise.
Due to drift errors in the device, where parameters can vary over the course of a day, we organized the experiments into separate groups, each completed at different times throughout the day.

\begin{figure*}[tb]
\begin{center}
\includegraphics[clip = true, width =2\columnwidth]{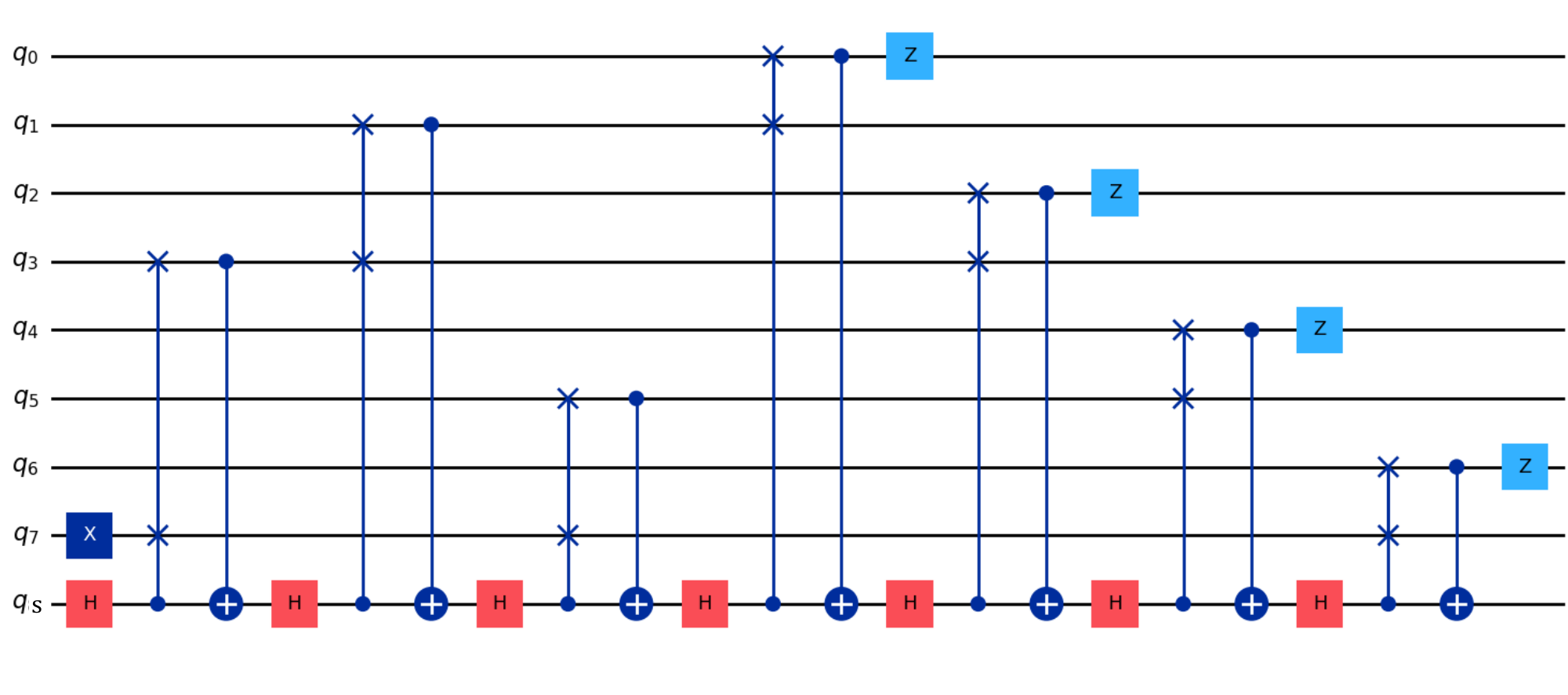}
\caption{\label{fig:PreS1} (color online) Circuit diagram for preparing the first scar state in the experiment.}
\end{center}
\end{figure*}

Several methods exist for generating exact scar states \cite{Gustafson2023,cenedese2024shallow}. For producing the first exact scar state, we employ a more convenient approach.
The procedure for generating the first scar state is outlined in Fig. \ref{fig:PreS1}. All the qubits are initialized to $|0\rangle$. The last qubit $q_{n-1}$ is flipped to $|1\rangle$ using a not gate (X). An equal superposition is created on the ancilla qubit with a Hadamard gate (H). This is followed by a controlled-SWAP gate, which generates a superposition of spins at different positions $\left(|0_{\rm a}\rangle \otimes | 100 \cdots \rangle + |1_{\rm a}\rangle \otimes | \cdots 010 \cdots \rangle\right)/\sqrt{2}$. Next, a CNOT gate is applied to restore the ancilla qubit, resulting in $|0_{\rm a}\rangle \otimes \left(| 100 \cdots \rangle +  | \cdots 010 \cdots \rangle\right)/\sqrt{2}$. Uniform distribution along the chain requires $O( n )$ equal divisions. Finally, a Z gate is applied to adjust the phase.

Experimental error mitigation based on fidelity is complex and will be discussed in detail below. We first measure the input state fidelity of the QCNN. Subsequently, we perform another experiment to prepare the same state and input it into the QCNN to obtain classification results. The data is processed using zero-error extrapolation \cite{Qin2022}, which allows us to estimate the net performance of the learning circuit.

The fidelity of the all-zero state can be measured directly, whereas measuring the fidelity of the first scar state is more challenging. To address this, we approximate that the fidelity of the prepared circuit is consistent across different states. Therefore, we use the fidelity of the easily-measured all-zero state as a proxy for the fidelity of the first scar state.

The error in single-qubit gates is significantly smaller than that in two-qubit gates; therefore, we neglect the errors in single-qubit gates. The unitary operation of the preparation circuit is the $U$ that satisfies $|\mathscr{S}_1\rangle=U|\mathscr{S}_0\rangle$. Additionally, we observe that $|\mathscr{S}_0\rangle=H_{\rm s}UX_{n-1}|\mathscr{S}_0\rangle$.
The fidelity of the output can be expressed as
\begin{eqnarray}
    {\rm Tr}\left\{ |\mathscr{S}_1\rangle\langle\mathscr{S}_1| \rho_\varepsilon \right\}&=&{\rm Tr}\large\{ U|\mathscr{S}_0\rangle\langle\mathscr{S}_0|U^\dagger U_\varepsilon |\mathscr{S}_0\rangle\langle\mathscr{S}_0|U_\varepsilon^\dagger  \large\} ~,\nonumber\\
\label{eq:UUap}
\end{eqnarray}
where the subscript $\varepsilon$ means the operation with error.
In a rough approximation, we make the replacement that
\begin{eqnarray}
   X_{n-1} U^\dagger U_\varepsilon X_{n-1}\approx U^\dagger U_\varepsilon ~.
\end{eqnarray}
This equation represents a zeroth-order approximation, which is exact in the absence of errors.
By substituting it into Eq. (\ref{eq:UUap}), we can achieve the transformation
\begin{eqnarray}
    &&{\rm Tr}\left\{ |\mathscr{S}_1\rangle\langle\mathscr{S}_1| \rho_\varepsilon \right\} \nonumber\\
&\approx&{\rm Tr}\large\{ UX_{n-1}|\mathscr{S}_0\rangle\langle\mathscr{S}_0|X_{n-1}U^\dagger \nonumber\\
    &&U_\varepsilon X_{n-1} |\mathscr{S}_0\rangle\langle\mathscr{S}_0|X_{n-1} U_\varepsilon^\dagger  \large\} \nonumber\\
    &\approx&{\rm Tr}\large\{ H_{\rm s}UX_{n-1}|\mathscr{S}_0\rangle\langle\mathscr{S}_0|X_{n-1}U^\dagger H_{\rm s} \nonumber\\
    &&H_{\rm s}U_\varepsilon X_{n-1} |\mathscr{S}_0\rangle\langle\mathscr{S}_0|X_{n-1} U_\varepsilon^\dagger H_{\rm s} \large\} \nonumber\\
    &\approx&{\rm Tr}\large\{ |\mathscr{S}_0\rangle\langle\mathscr{S}_0|H_{\rm s}U_\varepsilon X_{n-1} |\mathscr{S}_0\rangle\langle\mathscr{S}_0|X_{n-1} U_\varepsilon^\dagger H_{\rm s} \large\} ~. \nonumber\\
\end{eqnarray}
Instead of measuring the fidelity of the state $|\mathscr{S}_1\rangle$ after applying the circuit $U_\varepsilon$, we measure the fidelity of the state $|\mathscr{S}_0\rangle$ after applying the circuit $H_{\rm s}U_\varepsilon X_{n-1}$. This approach makes the experimental estimation more feasible.

\section{PXP Model Under Perturbations \label{App:Pert}}

\begin{figure}[tb]
\begin{center}
\includegraphics[clip = true, width =\columnwidth]{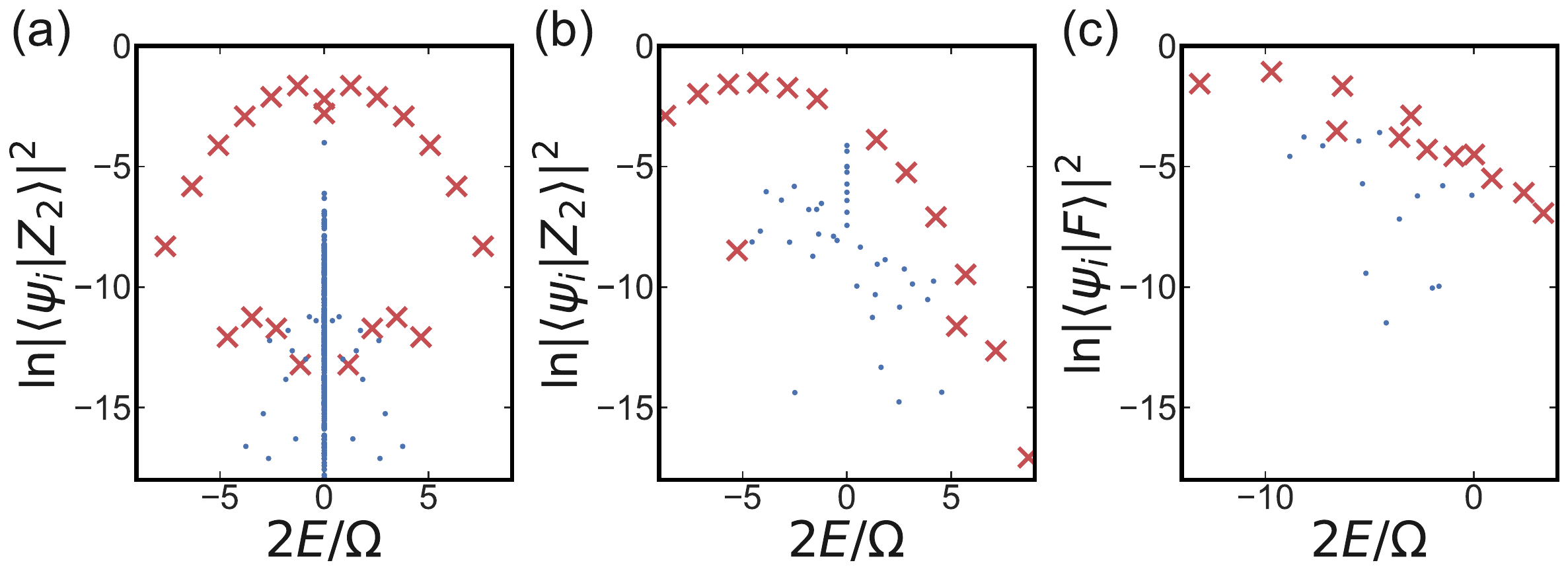}
\caption{\label{fig:PXPPer} (color online) (a) The eigenstates of $H_{\rm PXPZ}$ with perturbation strength $\lambda=0.05$. (b) The eigenstates of $H_{\rm sta}$ with perturbation strength $\lambda=0.35$. (c) The eigenstates of $H_{\rm Z}$ with perturbation strength $\lambda=0.7$.}
\end{center}
\end{figure}

The perturbative PXP models have garnered significant interest in the community due to their ability to alter both the spectral distribution and quench dynamics. In this section, we numerically evaluate the performance of the QCNN with continua boundary condition under Hamiltonian perturbations. Specifically, we consider three distinct types of perturbations.

Scars can be enhanced by incorporating higher-order corrections into the Hamiltonian, which in turn strengthens the revival. This modification can be expressed as \cite{Choi2019,Khemani2019}
\begin{eqnarray}
    H_{\rm PXPZ}&&=H_{\rm PXP}
    \nonumber
    \\
    &&-\lambda\sum_{i=0}^n \left( \sigma_{i-2}^zP_{i-1}^0\sigma_i^x P_{i+1}^0 + P_{i-1}^0\sigma_i^x P_{i+1}^0\sigma_{i+2}^z\right) ~, \nonumber\\
\end{eqnarray}
where $\lambda$ is the perturbation strength.

A staggered magnetic field induces the confinement of excitations, effectively isolating the antiferromagnetic-like states from other states. This separation also enhances the revival strength. The corresponding Hamiltonian is \cite{Desaules2024}
\begin{eqnarray}
    H_{\rm sta}=H_{\rm PXP}+\lambda\sum_{i=0}^n (-1)^i\sigma_i^z ~.
\end{eqnarray}

In the presence of a chemical potential, the reference state shifts to the ferromagnetic state $|F\rangle=|0\rangle^{\otimes n}$. The corresponding Hamiltonian is \cite{Daniel2023,Su2023}
\begin{eqnarray}
    H_{\rm Z}=H_{\rm PXP}+\lambda\sum_{i=0}^n \sigma_i^z ~.
\end{eqnarray}

We train the QCNN using the scar states in the top band for each perturbative case. In all cases, the QCNN identifies most of the scar states in the top band, as shown in Fig. \ref{fig:PXPPer}. For the revival-enhanced cases in Fig. \ref{fig:PXPPer} (a) and (b), fewer additional states are detected, suggesting that fewer scar-like states exist in these perturbative systems. This indicates that the dispersion is suppressed.
For the reference state switch case shown in Fig. \ref{fig:PXPPer} (c), the QCNN also identifies states in the second band below the top band, implying that these states share similar properties with the scar states. A more detailed and in-depth investigation is needed in future studies.

\end{document}